\documentclass[reprint,superscriptaddress,amsmath,amssymb,aps,prl,nofootinbib]{revtex4-2}
\usepackage{graphicx}
\usepackage{dcolumn}
\usepackage[colorlinks=true,linkcolor=black,citecolor=black]{hyperref}
\usepackage{siunitx}
\usepackage{textgreek}
\usepackage{wasysym}
\usepackage[super]{nth}
\usepackage[version=4]{mhchem}
\usepackage{stackengine}
\usepackage[T1]{fontenc}

\newcommand{\sth}{$\theta_{14}$}
\newcommand{\sthh}{$\sin^{2}(2\theta_{14})$}
\newcommand{\dms}{$\Delta m_{41}^{2}$}
\newcommand{\urm}{\ce{^{235}U}}
\newcommand{\urmm}{\ce{^{238}U}}
\newcommand{\rdn}{\ce{^{222}Rn}}
\newcommand{\overbar}[1]{\mkern 1.0mu\overline{\mkern-1.0mu#1\mkern-1.0mu}\mkern 1.0mu}

\begin{document}

\preprint{APS/123-QED}

\title{Search for Very-Short-Baseline Oscillations of\\Reactor Antineutrinos with the SoLid Detector}

\author{Y.~Abreu}\affiliation{Universiteit Antwerpen, Antwerpen, Belgium}
\author{Y.~Amhis}\affiliation{IJCLab, Univ Paris-Sud, CNRS/IN2P3, Université Paris-Saclay, Orsay, France}
\author{L.~Arnold}\affiliation{University of Bristol, Bristol, United Kingdom}
%\author{G.~Barber}\affiliation{Imperial College London, Department of Physics, London, United Kingdom}
\author{W.~Beaumont}\affiliation{Universiteit Antwerpen, Antwerpen, Belgium}
%\author{S.~Binet}\affiliation{Université Clermont Auvergne, CNRS/IN2P3, LPCA, Clermont-Ferrand, France}
\author{I.~Bolognino}\affiliation{Department of Physics, The University of Adelaide, Adelaide, SA 5005, Australia}
\author{M.~Bongrand}\affiliation{IJCLab, Univ Paris-Sud, CNRS/IN2P3, Université Paris-Saclay, Orsay, France}
%\author{J.~Borg}\affiliation{Imperial College London, Department of Physics, London, United Kingdom}
\author{D.~Boursette}\affiliation{IJCLab, Univ Paris-Sud, CNRS/IN2P3, Université Paris-Saclay, Orsay, France}
\author{V.~Buridon}\affiliation{Normandie Univ., ENSICAEN, UNICAEN, CNRS/IN2P3, LPC Caen, Caen, France}
%\author{B.C.~Castle}\affiliation{University of Oxford, Oxford, United Kingdom}
\author{H.~Chanal}\affiliation{Université Clermont Auvergne, CNRS/IN2P3, LPCA, Clermont-Ferrand, France}
%\author{K.~Clark}\affiliation{University of Bristol, Bristol, United Kingdom}
\author{B.~Coupé}\affiliation{SCK CEN, Belgian Nuclear Research Centre, Mol, Belgium}
\author{P.~Crochet}\affiliation{Université Clermont Auvergne, CNRS/IN2P3, LPCA, Clermont-Ferrand, France}
\author{D.~Cussans}\affiliation{University of Bristol, Bristol, United Kingdom}
\author{J.~D'Hondt}\affiliation{Vrije Universiteit Brussel, Brussel, Belgium}
\author{D.~Durand}\affiliation{Normandie Univ., ENSICAEN, UNICAEN, CNRS/IN2P3, LPC Caen, Caen, France}
%\author{T.~Durkin}\affiliation{STFC, Rutherford Appleton Laboratory, Harwell Oxford, and Daresbury Laboratory, Warrington, United Kingdom}
\author{M.~Fallot}\affiliation{SUBATECH, Nantes Université, IMT Atlantique, CNRS/IN2P3, Nantes, France}

\author{D.~Galbinski}
\email{galbinski@lpccaen.in2p3.fr}
\affiliation{Normandie Univ., ENSICAEN, UNICAEN, CNRS/IN2P3, LPC Caen, Caen, France}

\author{S.~Gallego}\affiliation{Johannes Gutenberg University of Mainz, Institute of Physics, Mainz}
\author{L.~Ghys}\affiliation{SCK CEN, Belgian Nuclear Research Centre, Mol, Belgium}
\author{L.~Giot}\affiliation{SUBATECH, Nantes Université, IMT Atlantique, CNRS/IN2P3, Nantes, France}
\author{K.~Graves}\affiliation{Imperial College London, Department of Physics, London, United Kingdom}
\author{B.~Guillon}\affiliation{Normandie Univ., ENSICAEN, UNICAEN, CNRS/IN2P3, LPC Caen, Caen, France}
\author{S.~Hayashida}\affiliation{Kings College London, London, United Kingdom}
\author{D.~Henaff}\affiliation{SUBATECH, Nantes Université, IMT Atlantique, CNRS/IN2P3, Nantes, France}
\author{B.~Hosseini}\affiliation{Imperial College London, Department of Physics, London, United Kingdom}
%\author{S.~Jenzer}\affiliation{IJCLab, Univ Paris-Sud, CNRS/IN2P3, Université Paris-Saclay, Orsay, France}
\author{S.~Kalcheva}\affiliation{SCK CEN, Belgian Nuclear Research Centre, Mol, Belgium}
\author{L. N.~Kalousis}\affiliation{Vrije Universiteit Brussel, Brussel, Belgium}
\author{R.~Keloth}\affiliation{Vrije Universiteit Brussel, Brussel, Belgium}
\author{L.~Koch}\affiliation{Johannes Gutenberg University of Mainz, Institute of Physics, Mainz}
\author{M.~Labare}\affiliation{Universiteit Gent, Gent, Belgium}

\author{G.~Lehaut}
\email{lehaut@lpccaen.in2p3.fr}
\affiliation{Normandie Univ., ENSICAEN, UNICAEN, CNRS/IN2P3, LPC Caen, Caen, France}

\author{S.~Manley}\affiliation{University of Bristol, Bristol, United Kingdom}
\author{L.~Manzanillas}\affiliation{IJCLab, Univ Paris-Sud, CNRS/IN2P3, Université Paris-Saclay, Orsay, France}
\author{J.~Mermans}\affiliation{SCK CEN, Belgian Nuclear Research Centre, Mol, Belgium}
\author{I.~Michiels}\affiliation{Universiteit Gent, Gent, Belgium}
\author{S.~Monteil}\affiliation{Université Clermont Auvergne, CNRS/IN2P3, LPCA, Clermont-Ferrand, France}
\author{C.~Moortgat}\affiliation{Universiteit Gent, Gent, Belgium}
\author{D.~Newbold}\affiliation{STFC Rutherford Appleton Laboratory, Didcot, United Kingdom}
\author{V.~Pestel}\affiliation{Normandie Univ., ENSICAEN, UNICAEN, CNRS/IN2P3, LPC Caen, Caen, France}
\author{K.~Petridis}\affiliation{University of Bristol, Bristol, United Kingdom}
\author{I.~Piñera}\affiliation{Université Clermont Auvergne, CNRS/IN2P3, LPCA, Clermont-Ferrand, France}
%\author{L.~Popescu}\affiliation{SCK CEN, Belgian Nuclear Research Centre, Mol, Belgium}
\author{A.~de Roeck}\affiliation{Universiteit Antwerpen, Antwerpen, Belgium}
\author{N.~Roy}\affiliation{IJCLab, Univ Paris-Sud, CNRS/IN2P3, Université Paris-Saclay, Orsay, France}
\author{D.~Ryckbosch}\affiliation{Universiteit Gent, Gent, Belgium}
\author{N.~Ryder}\affiliation{University of Oxford, Oxford, United Kingdom}
\author{D.~Saunders}\affiliation{University of Bristol, Bristol, United Kingdom}
\author{M. H.~Schune}\affiliation{IJCLab, Univ Paris-Sud, CNRS/IN2P3, Université Paris-Saclay, Orsay, France}
\author{M.~Settimo}\affiliation{SUBATECH, Nantes Université, IMT Atlantique, CNRS/IN2P3, Nantes, France}
\author{H.~Rejeb Sfar}\affiliation{Universiteit Antwerpen, Antwerpen, Belgium}
\author{L.~Simard}\affiliation{IJCLab, Univ Paris-Sud, CNRS/IN2P3, Université Paris-Saclay, Orsay, France}

\author{A.~Vacheret}
\email{vacheret@lpccaen.in2p3.fr}
\affiliation{Normandie Univ., ENSICAEN, UNICAEN, CNRS/IN2P3, LPC Caen, Caen, France}

\author{S.~Van Dyck}\affiliation{SCK CEN, Belgian Nuclear Research Centre, Mol, Belgium}
\author{P.~Van Mulders}\affiliation{Vrije Universiteit Brussel, Brussel, Belgium}
\author{N.~Van Remortel}\affiliation{Universiteit Antwerpen, Antwerpen, Belgium}
\author{G.~Vandierendonck}\affiliation{Universiteit Gent, Gent, Belgium}
\author{S.~Vercaemer}\affiliation{Universiteit Antwerpen, Antwerpen, Belgium}
\author{M.~Verstraeten}\affiliation{Ecole Royale Militaire/Koninklijke Militaire School, Plasma Physics Laboratory, Brussel, Belgium}
\author{B.~Viaud}\affiliation{SUBATECH, Nantes Université, IMT Atlantique, CNRS/IN2P3, Nantes, France}
\author{A.~Weber}\affiliation{Fermi National Accelerator Laboratory, Batavia, USA}\affiliation{Johannes Gutenberg University of Mainz, Institute of Physics, Mainz}
\author{M.~Yeresko}\affiliation{Université Clermont Auvergne, CNRS/IN2P3, LPCA, Clermont-Ferrand, France}
\author{F.~Yermia}\affiliation{SUBATECH, Nantes Université, IMT Atlantique, CNRS/IN2P3, Nantes, France}
\collaboration{SoLid Collaboration}

\date{\today}

\begin{abstract}

In this letter we report the first scientific result based on antineutrinos emitted from the BR2 reactor at SCK CEN. The SoLid experiment uses a novel type of highly granular detector whose basic detection unit combines two scintillators, PVT and $^6$LiF:ZnS(Ag), to measure antineutrinos via their inverse-beta-decay products. An advantage of PVT is its highly linear response as a function of deposited particle energy. The full-scale detector comprises \num{12800} voxels and operates over a very short 6.3--8.9\,m baseline from the reactor core. The detector segmentation and its 3D imaging capabilities facilitate the extraction of the positron energy from the rest of the visible energy, allowing the latter to be utilised for signal-background discrimination. We present a result based on 280 reactor-on days (55\,\si{\mega\watt} mean power) and 172 reactor-off days, respectively, of live data-taking. A total of $\num{29479} \pm 603$ (stat.\@) antineutrino candidates have been selected, corresponding to an average rate of 105 events per day and a signal-to-background ratio of 0.27. A search for disappearance of antineutrinos to a sterile state has been conducted using complementary model-dependent frequentist and Bayesian fits, providing constraints on the allowed region of the Reactor Antineutrino Anomaly.

\end{abstract}

\maketitle

%%%%%%%%%%%%%%%%%%%%
%%% INTRODUCTION %%%
%%%%%%%%%%%%%%%%%%%%

The observation of the Reactor Antineutrino Anomaly (RAA) in 2011 \cite{Mention2011} initiated a flurry of activity in very-short-baseline neutrino physics. It followed from updated reactor-antineutrino flux predictions, known collectively as the Huber-Mueller (HM) model \cite{Mueller2011, Huber2011}, which retroactively revealed a deficit in the electron-antineutrino rate measured by numerous short-baseline experiments. Since then, antineutrino flux predictions from both the conversion \cite{Hayen2019,Li2019,Kopeikin2021} and ``ab initio'' summation \cite{Estienne2019,Perisse2023,Letourneau2023} methods have evolved substantially. The statistical significance of the RAA---initially 2.6\textsigma{}---varies greatly depending on the reactor model \cite{Giunti2022}. The current state-of-the-art calculations tentatively point to \urm{} as the principle source of the rate deficit, likely induced by biases in the ILL data \cite{Schreckenbach1981,Feilitzsch1982,Schreckenbach1985} on which the HM model is based, or poorly understood uncertainties \cite{Hayes2016} and unreliable nuclear databases \cite{Sonzogni2016}. This is supported by results from Daya Bay \cite{An2017,An2023} and RENO \cite{Bak2019} on the correlation between antineutrino yield and fuel composition, which suggest that adjusting either the predicted \urm{} rate, or all fissile isotopes equally, could resolve the RAA.

Alternatively, the RAA could be explained by disappearance of reactor antineutrinos to an eV-scale \textit{sterile} state. Sterile neutrinos would not participate in any Standard Model interaction, making them accessible to experiments only by mixing with ordinary \textit{active} neutrinos. Assuming one new neutrino mass eigenstate (3+1 model), the two-flavour antineutrino survival probability as a function of energy ($E$) and distance ($L$) is given by:

\begin{equation}\label{eq:survival}
    P = 1 - \sin^2(2\theta_{14})\sin^2\left(1.27\:\Delta m^2_{41}[\si{\electronvolt\squared}]\:\frac{L[\si{\metre}]}{E[\si{\mega\electronvolt}]} \right) \:,
\end{equation}

where \sth{} and \dms{} are the mixing angle and mass splitting, respectively, between the active and sterile states. Given the typical energy of reactor antineutrinos (1--10\,\si{\mega\electronvolt}), the mass splitting implied by the RAA means that oscillations occurring over $\mathcal{O}$(1)\,m distances can be searched for using detectors placed in extreme proximity to reactor cores. The sterile-neutrino hypothesis of the RAA has been thoroughly tested at both research \cite{prospect2,stereo2,Serebrov2017} and commercial \cite{Alekseev2018,Ko2017} reactors. No evidence supporting the existence of sterile neutrinos has been found, with the exception of the unconfirmed Neutrino-4 result \cite{Serebrov2021,badnu4a} which is in strong tension with most limits. It is, however, consistent with the best-fit region of the Gallium anomaly \cite{Abazov1991,Hampel1998}, recently refreshed by BEST \cite{Barinov2022}. SoLid (Search for Oscillations with a Lithium-6 Detector) is another such very-short-baseline experiment \cite{Abreu2017,Abreu2018,Abreu2018v2,Abreu2019v2,Abreu2019,Abreu2021}, whose main goal is to search for disappearance of reactor antineutrinos.

%%%%%%%%%%%%%%%%%%%%%%%%
%%% SOLID EXPERIMENT %%%
%%%%%%%%%%%%%%%%%%%%%%%%

\begin{figure}
    \centering
    \includegraphics[width=1.0\linewidth]{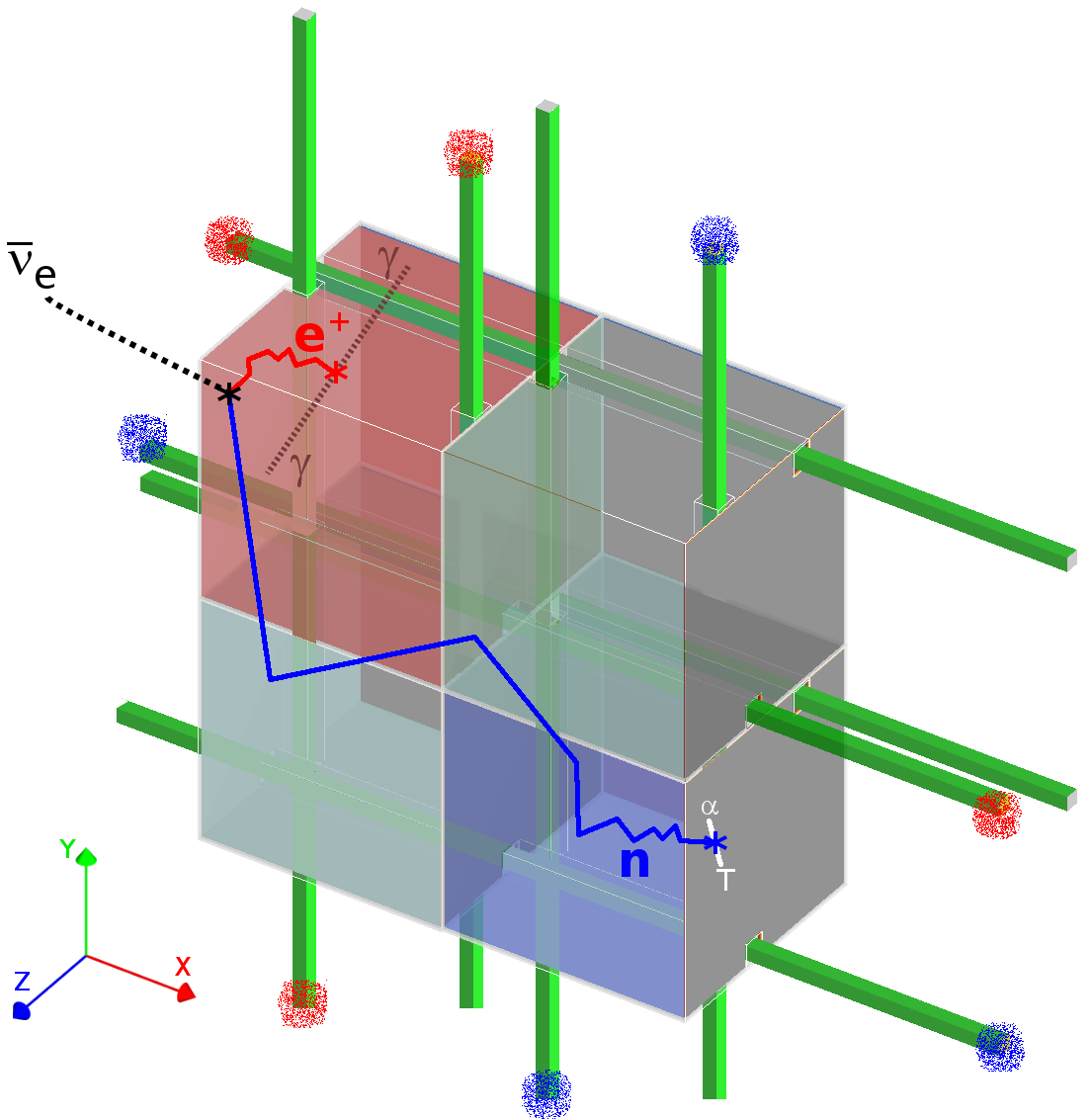}
    \caption{SoLid antineutrino detection principle. The positron and neutron from inverse beta decay create two scintillation signals, separated by a characteristic time delay ($\Delta t \sim 62$\,\si{\micro\second}) during which the neutron thermalises over $\mathcal{O}$(10)\,cm distances.}
    \label{fig:detector}
\end{figure}

SoLid is operated next to the BR2 reactor at SCK CEN in Mol, Belgium. BR2 is an open-pool research reactor that burns highly enriched uranium (93.5\% \urm{}) at 40--100\,\si{\mega\watt} thermal power. The unique inclined-channel design results in a compact core ($\diameter$ \SI{50}{\centi\metre} $\times$ \SI{80}{\centi\metre}) for fission reactions. The reactor operates at power in cycles of around 30 days each, alternating with shut-down periods of comparable duration. The SoLid detector is based on a combination of two scintillators: 5\,cm polyvinyl toluene (PVT) cubes and $^6$LiF:ZnS(Ag) screens. Antineutrinos interact in the fiducial volume of the detector through inverse beta decay (IBD), ${\overbar{\nu}_{e} + p \rightarrow n + e^{+}}$, producing a positron and a neutron. PVT acts as a proton-rich target for the incoming antineutrino, scintillator for the positron, and moderator of the neutron. A schematic of the detection principle is displayed in Fig.~\ref{fig:detector}. Accompanying the positron signal are energy deposits from Compton scattering interactions of the \SI{511}{\kilo\electronvolt} \textgamma{}-rays emitted when the positron annihilates (${e^{+} e^{-} \rightarrow \gamma \gamma}$). The total prompt signal is referred to as the Electromagnetic Signal (ES). The $^6$LiF:ZnS(Ag) screens on two faces of each PVT cube are held in place by Tyvek wrapping, which also ensures optical isolation of the voxels. The neutron transport was optimised by tuning the number of $^6$LiF:ZnS(Ag) screens per cube, the thickness of the screens, and their ZnS-$^6$LiF mass ratio. After an average delay ($\Delta t$) of $\sim 62$\,\si{\micro\second}, thermalised neutrons are captured on $^6$Li via the reaction $\ce{^{6}Li + \text{n} \rightarrow ^{3}_{2}\text{He} + \alpha}$, whose products cause fluorescence in the ZnS(Ag). This is labelled the Nuclear Signal (NS). An ES-NS time-coincidence is the primary signature of IBD-like events \cite{Reines1953v2}.

The 1.6\,\si{\tonne} detector comprises \num{12800} cubes arranged into five mechanically independent modules of ten planes each, where a plane is a grid of $16 \times 16$ cubes. Each cube is crossed by four wavelength-shifting fibres in the directions perpendicular to the reactor-detector axis. One end of each fibre is capped by a Hamamatsu S12572-050P multi-pixel photon counter (MPPC) and the other by reflective foil. The detector is housed at BR2 in a light-proof, environmentally controlled shipping container to ensure consistent data-taking conditions. This is topped by passive \SI{50}{\centi\metre} HDPE shielding to add to the minimal 8\,m.w.e.\@ overburden of the BR2 building and a \SI{50}{\centi\metre} water wall on all sides of the container helps to thermalise cosmic neutrons and block reactor-induced gamma background. During normal data-taking, an FPGA-based trigger system is deployed \cite{Abreu2019}. For recording IBD-like events, the approach taken is to trigger on NS rather than ES waveforms which are dominated by gamma-background events. Upon firing of this trigger, a space-time window large enough to encapsulate the entire ES-NS event is read out and saved to disk.

The digitised signals from the readout are organised into discrete clusters based on the natural temporal and spatial correlations between waveforms from the same physics event, whilst neglecting dark counts via an amplitude threshold. Through-going muons deposit energy in the cubes along their path, leaving highly distinguishable tracks. Other clusters are tagged as either ES or NS by a simple integral-over-amplitude cut, which enhances the purity of the NS sample from the 20\% trigger-only output to over 99\%. It is then necessary to allocate the MPPC signals to the cubes from where they originated and assign them corresponding energies. Since physics events typically generate energy deposits that span multiple cubes and channels, this problem is non-trivial. Defining $a_{ij}$ as the projector of cube $j$ to MPPC $i$, the equation ${A E = p}$ must be solved for every plane in which energy was deposited, where $p$ is the vector of MPPC signals ($p_1, p_2,...,p_{64}$), $E$ is the list of unknown cube energies ($E_1,E_2,...,E_{256}$), and $A$ is a $64 \times 256$ matrix dubbed the \textit{system matrix}. Solutions are obtained using a custom algorithm called CCube, which combines Maximum-Likelihood Expectation-Maximisation \cite{next} with a simplified version of the Orthogonal Matching Pursuit algorithm \cite{omp}. An extensive description of CCube is provided in Ref.~\cite{Abreu2024}.

\begin{figure}
    \centering
    \includegraphics[width=1.0\linewidth]{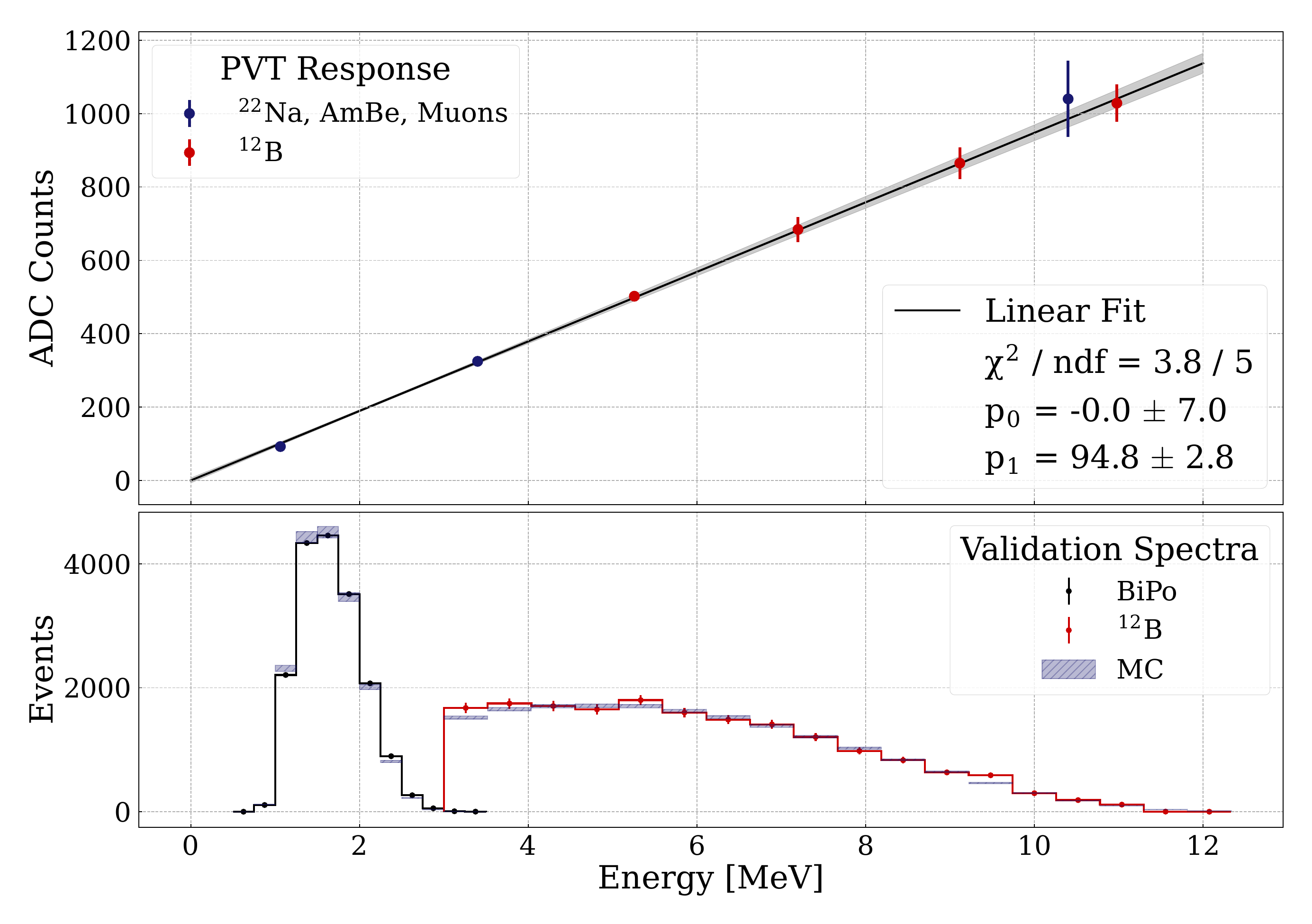}
    \caption{Top: Measured response as a function of energy of a subset of the PVT detector cubes, from September 2018 calibration data. The linear fit is derived from the blue points and constrained to intercept the vertical axis at zero; it is validated by the red points which align well with the fit. The grey band represents the error on the fitted gradient. All but the first \ce{^{12}B} point are incorporated in the $\chi^2$. Bottom: Data-MC comparisons of BiPo and boron-12 spectra.}
    \label{fig:linear}
\end{figure}

The system matrix parameterises vital information about the detector response---such as scintillation efficiency and photon attenuation---and provides a relative calibration. The responses of individual cubes can be equalised detector-wide by fitting the $\text{d}E/\text{d}x$ energy-loss distributions of horizontal muons (i.e.\@ those that cross only one cube per plane). Moreover, the fractions of light shared between cubes ($\mathcal{O}$(1)\% due to the fibre-slot clearance) are evaluated and factored into the matrix. Muons can also be used to estimate the detector energy scale, but the associated uncertainties are relatively large. The absolute light yield is determined more precisely using in-situ radioactive sources, \ce{^{22}Na} and \ce{AmBe}, by fitting the Compton edge profile of their emitted gamma radiation \cite{Abreu2021}. The corresponding energy resolution is approximately 16\%. Alternatively, the energy calibration requirements can be fulfilled by the monoenergetic 3.43\,\si{\mega\electronvolt} electron-positron pair produced by the 4.44\,\si{\mega\electronvolt} AmBe \textgamma{}-ray. This relatively low-rate signal ($\sim 30$ times less likely than Compton scattering) can be identified via the same techniques as employed for antineutrino events \cite{mike} and allows the light yield to be computed with a percent-level uncertainty. Validations of the energy scale and MC detector simulation are provided by the \textbeta{}-spectra of cosmogenic \ce{^{12}B} and \ce{_{83}^{214}Bi} (of BiPo background), as shown in Fig.~\ref{fig:linear} along with the measured PVT response. The high degree of linearity attained in the 1--11\,\si{\mega\electronvolt} region demonstrates how the SoLid technology complements similar liquid-scintillator-based experiments.

%%%%%%%%%%%%%%%%%%%%%%%
%%% THE DATASET %%%
%%%%%%%%%%%%%%%%%%%%%%%

The dataset on which this analysis is based spans data taken between \nth{11} April 2018 to \nth{2} July 2020. It comprises 13 reactor-on (ROn) cycles with a total live-time of 280.3 days, and a total reactor-off (ROff) live-time of 171.8 days. Detector stability was continuously monitored through low-level DAQ parameters including MPPC pedestal and gain, which were regularly re-equalised across all channels. The dataset has been filtered through stringent data-quality criteria based on various reconstructed quantities and restrictions on environmental conditions such as temperature and humidity. Periodic calibration campaigns with AmBe and \ce{^{22}Na} were used to track the time-evolution of the neutron reconstruction efficiency and energy scale, respectively. The former remained stable across the detector planes to within 1\%, while both exhibited drifts of a few percent per year caused by degradation of the scintillating materials. The muon-derived system matrix for the ES calibration was updated roughly every ten data-taking days to accommodate shifts in detector response, particularly the light yield. For the analysis, the detector system and surrounding environment are accurately implemented in \textsc{Geant4} \cite{geant4}, including all materials that could influence particle interactions within the detector volume. A second software framework then applies readout effects to the simulated energy deposits according to a system matrix. Fission yields are simulated using an MCNPX \cite{mcnp} representation of the BR2 reactor core coupled with MCNPX/CINDER90 burn-up code; off-equilibrium effects are taken into account in MURE \cite{mure}. Conversion antineutrino spectra are then constructed by linking the fission vertices to the ILL data and summation spectra by aggregating the \textbeta{}-decay branches of all relevant isotopes.

%%%%%%%%%%%%%%%%%%%%
%%% IBD ANALYSIS %%%
%%%%%%%%%%%%%%%%%%%%

SoLid faces three IBD-like backgrounds: fast neutrons, from cosmic-ray-induced spallation of atmospheric nuclei and materials surrounding the detector; radioactive decays, both from airborne \rdn{} and \urmm{} contamination in the $^6$LiF:ZnS(Ag) screens; and a low rate of ``accidental'' events from time-coincidences of uncorrelated signals. Fast neutrons cause proton recoils (ES) prior to their capture on $^6$LiF (NS) after thermalisation. The radioactive background, BiPo, produces ES-NS coincidences from radiative \textbeta{}-decay of \ce{_{83}^{214}Bi}, followed by emission of an \textalpha{}-particle from \ce{_{84}^{214}Po}---which has a decay time of \SI{235.8}{\micro\second}---generating scintillation light in the ZnS(Ag). The exceedingly high initial background rate is suppressed by application of the selection cuts outlined in Tab.~\ref{tab:cuts}. Cuts 1--6 are based on low-level event-reconstruction variables. In particular, the NS-ES displacement cuts reduce the number of accidental events to an almost negligible level. The muon veto decreases the number of fast-neutron events created after muon spallation in materials in and around the detector; all NS clusters that follow within \SI{200}{\micro\second} of a muon track are rejected.

\begin{table}
\renewcommand{\arraystretch}{1}
   \centering
   \begin{tabular}{rl|c}
     \hline \hline
      & {\bf Description} & {\bf Applied Limits} \\ \hline
     1) & NS-ES Time Diff. & $1$\,\si{\micro\second} $\leq \Delta t \leq 141$\,\si{\micro\second} \\
     2) &  NS-ES Distance & $1 \leq \Delta R \leq 4$ \\
     3) &  NS-ES Displacement & $-3 \leq \Delta X \leq 3$ \\
     4) &  \textquotedbl & $-3 \leq \Delta Y \leq 3$ \\
     5) &  \textquotedbl & $-2 \leq \Delta Z \leq 3$ \\
     6) &  Muon Veto & $t_{\text{nearest muon}} \geq 200$\,\si{\micro\second}\\
     7) &  Prompt Energy & $1$\,\si{\mega\electronvolt} $\leq E_{\text{MEC}} \leq 6$\,\si{\mega\electronvolt} \\
     8) &  Gamma Multiplicity & $N_{\text{gammas}} \geq 1$ \\\hline
     9) &  CNN for NS signal & $0.70 \leq$ Score $\leq 1.00$ \\
     10) &  BDT for ES signal & $0.73 \leq$ Score $\leq 1.00$ \\\hline \hline
   \end{tabular}
   \caption{Selection cuts of the IBD analysis.}
   \label{tab:cuts}
\end{table}

For cuts 7 and 8, the detector granularity is exploited to isolate the positron energy, which carries the antineutrino spectrum information, and reconstruct the annihilation photons ($\gamma_{\text{a}}$). Simulations of antineutrino events (IBD MC) indicate that the most energetic cube (MEC) in the ES cluster contains on average 91.5\% and 12.5\% of the deposited positron and $\gamma_{\text{a}}$ energy, respectively. The positron is confined to a single voxel for around 78\% of IBD MC events. In other cases, the majority of the remaining positron energy is located in cubes directly adjacent to the MEC, which can also contain significant quantities of $\gamma_{\text{a}}$ energy. As such, the positron energy is defined as the energy of the MEC plus any bordering cubes that contain at least 20\% of the total ES energy (${E_{\text{cube}} / E_{\text{tot}} \geq 0.2}$). This estimator contains 96.7\% and 13.8\% of positron and $\gamma_{\text{a}}$ energy on average. Furthermore, low-energy (${E_{\text{cube}} / E_{\text{tot}} < 0.2}$) deposits in cubes not bordering the MEC are dominated by contributions from the annihilation gammas. Capitalising on the anti-parallel emission directions of the two photons, detached clusters of gamma deposits are searched for in opposing half-spheres around the MEC. The ``gamma multiplicity'' variable can thus take values 0, 1, or 2; background-rich 0-\textgamma{} events are discarded.

\begin{figure}
    \centering
    \includegraphics[width=1.0\linewidth]{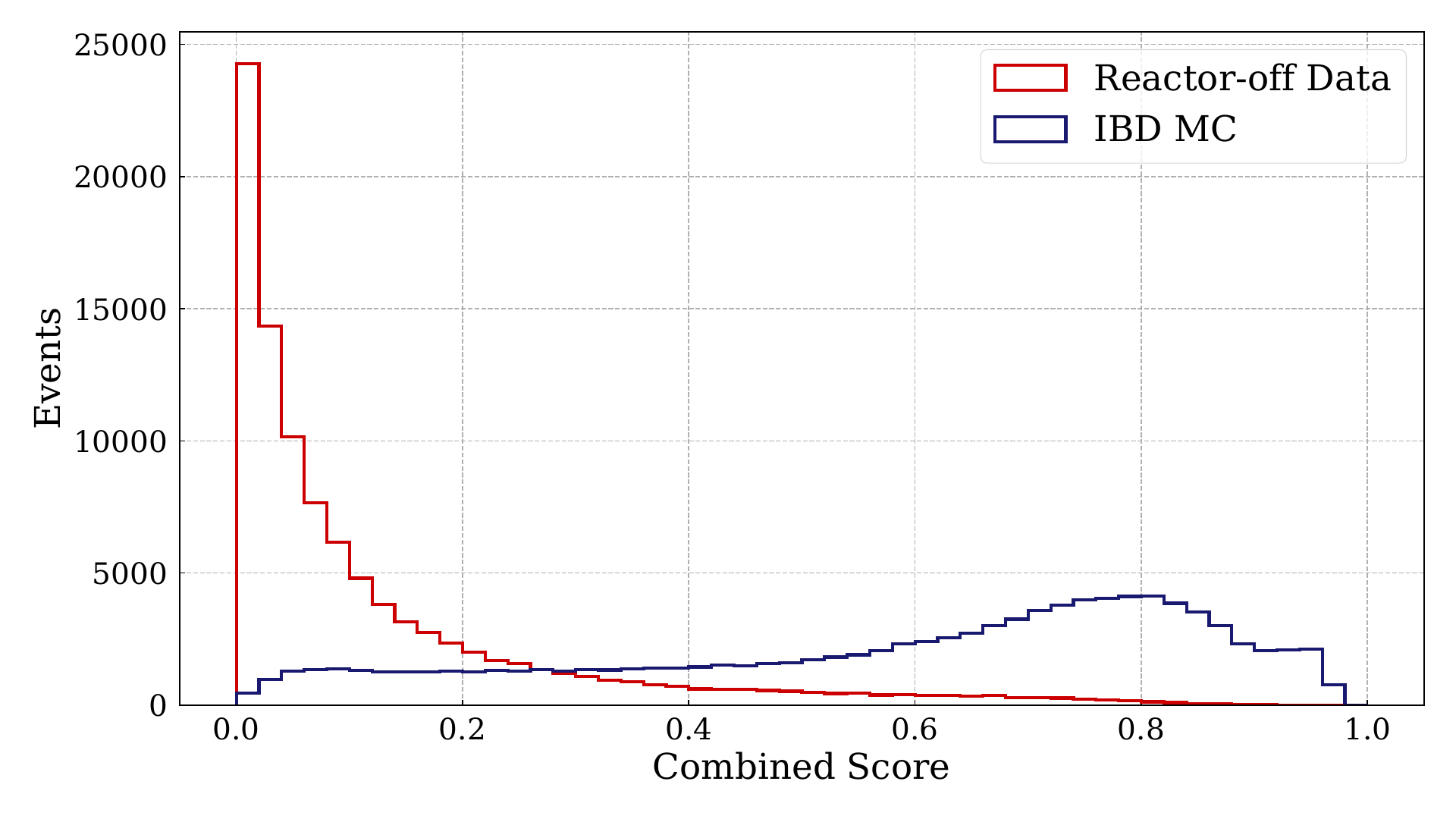}
    \caption{Separation between signal (IBD MC) and background (reactor-off) events achieved by the BDT and CNN models, expressed as the product of their two output scores.}
    \label{fig:ml}
\end{figure}

The final two cuts are applied on the outputs of machine-learning models. A one-dimensional Convolutional Neural Net (CNN) has been trained on pure data samples to distinguish between neutron- and alpha-induced NS waveforms using only pulse shape information. The CNN architecture is identical to the one described in Ref.~\cite{Griffiths2020} and it achieves 80\% neutron efficiency for rejection of 95\% of alpha waveforms. The second model is a Boosted Decision Tree (BDT) made with XGBoost \cite{xgboost}, primarily designed to discriminate between the ES clusters of IBD events and fast neutrons. It is trained on a dataset containing reactor-off events (background, labelled 0) and IBD MC events (signal, labelled 1) in equal proportions, after having first applied selection cuts 1--9. The twenty BDT features mainly comprise topological variables (e.g.\@ from the gamma reconstruction) that have no significant correlation with the positron energy or Z-position. Discrimination is also provided by the spatial-displacement parameters in Tab.~\ref{tab:cuts} and the event coordinates in the X--Y plane. The signal-background separation granted by these models is illustrated in Fig.~\ref{fig:ml}. Overall, cuts 1--9 remove 61\% of signal events and 98\% of background events; the BDT cut then removes an additional 39\% and 90\%, respectively. A comprehensive summary of the antineutrino selection may be found in Ref.~\cite{Galbinski2024}.

%%%%%%%%%%%%%%%%%%%%%%%%%%%%%%
%%% BACKGROUND SUBTRACTION %%%
%%%%%%%%%%%%%%%%%%%%%%%%%%%%%%

\begin{figure*}
    \centering
    \includegraphics[width=1.0\linewidth]{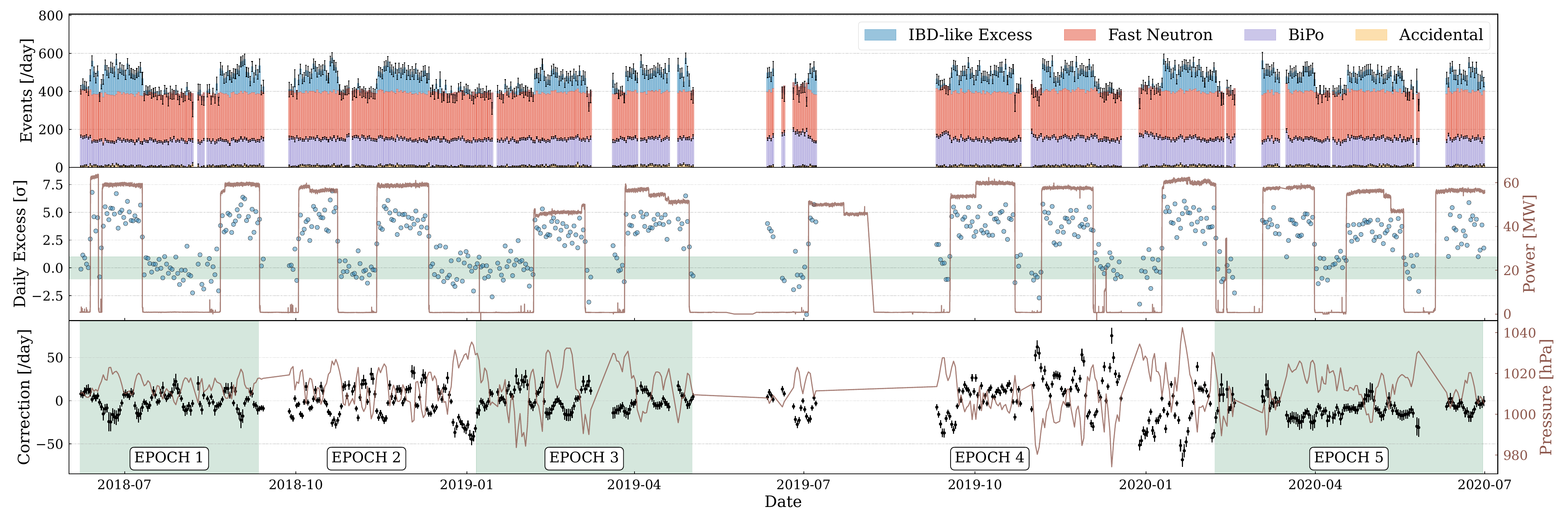}
    \caption{Overview of the SoLid dataset in terms of daily variations of different event types and atmospheric-pressure corrections.}
    \label{fig:test}
\end{figure*}

The oscillation fits are performed on a background-subtracted IBD spectrum. Subtraction of accidentals and BiPo is realised by estimating their respective rates and spectrum shapes in dedicated sideband regions of parameter space. For example, BiPo is characterised in a high-purity sideband that is nearly identical to the ``signal region'' (as defined by the Tab.~\ref{tab:cuts} selections) but which occupies a different region of $\Delta t$-CNN space. An equivalent region exists for isolating accidental background. The relative rates of different event types in a given sample are deduced from a triple-exponential fit on the $\Delta t$ distribution, adopting the known ES-NS time constants of each. To subtract fast-neutron events, it is essential to account for their dependence on atmospheric pressure. This is done by applying a daily correction computed from a linear fit of the fast-neutron rate $N^{\text{ROff}}_{\text{Signal}-\text{Acc}-\text{BiPo}}$ (accidental- and BiPo-subtracted reactor-off data) as a function of pressure $P$, both relative to their respective means over a set period:

\begin{equation}\label{eq:press_model}
\begin{aligned}
    f(P,\overbar{P}) = &\:\: N^{\text{ROff}}_{\text{Signal}-\text{Acc}-\text{BiPo}} - \overbar{N}^{\text{ROff}}_{\text{Signal}-\text{Acc}-\text{BiPo}} \\ = &\:\: p_0 + p_1 (P - \overbar{P}) \:,
\end{aligned}
\end{equation}

\noindent where $p_0$ and $p_1$ are the parameters of the fit. In practice, the robustness of the subtraction is improved by splitting the dataset into five independent ``epochs'', each with its own pressure fit $f_e$. The correction for a given day (time $t$) is thus determined by the composite model $F(t,P,\overbar{P}) = \sum_{e} \delta(t,e)\: f_e(P,\overbar{P})$, where the Kronecker-like function $\delta$ ensures the correct epoch. Denoting the number of events in a particular parameter-space region as $N$ and using a tilde to signify a pressure-corrected quantity, the number of IBD events can be expressed as:

\begin{widetext}
\begin{equation}
\begin{aligned}
    N_{\text{IBD}} = &\:\: \tilde{N}^{\text{ROn}}_{\text{Signal}-\text{Acc}-\text{BiPo}} - \frac{t_{\text{ROn}}}{t_{\text{ROff}}} \cdot \tilde{N}^{\text{ROff}}_{\text{Signal}-\text{Acc}-\text{BiPo}} \\ = &\:\: (N^{\text{ROn}}_{\text{Signal}} - N^{\text{ROn}}_{\text{Signal}\:\text{Acc}}) - \beta^{\text{ROn}} \cdot (N^{\text{ROn}}_{\text{BiPo}} - N^{\text{ROn}}_{\text{BiPo}\:\text{Acc}}) - \sum\limits_{i}^{I} F(t_i,P_i,\overbar{P}_{\text{ROn}}) \\ - &\: \frac{t_{\text{ROn}}}{t_{\text{ROff}}} \cdot [(N^{\text{ROff}}_{\text{Signal}} - N^{\text{ROff}}_{\text{Signal}\:\text{Acc}}) - \beta^{\text{ROff}} \cdot (N^{\text{ROff}}_{\text{BiPo}} - N^{\text{ROff}}_{\text{BiPo}\:\text{Acc}}) - \sum\limits_{j}^{J} F(t_j,P_j,\overbar{P}_{\text{ROff}})] \:,
\end{aligned}
\end{equation}
\end{widetext}

\noindent where $N_{\text{Signal}}$ and $N_{\text{BiPo}}$ refer to the number of events in the signal region and BiPo sideband, respectively, with $\beta$ correcting for the BiPo-rate difference between these two regions, and $t_{\text{ROn}} = \sum\limits t_i$ ($t_{\text{ROff}} = \sum\limits t_j$) is the total reactor-on (reactor-off) live-time corresponding to $I$ ($J$) data-taking days. Finally, the dataset is bisected after epoch 3 to create two halves (of roughly equal size in ROn data) for which the subtractions are performed using separate ROff data. The resulting spectra are then combined. This procedure was found to mitigate the need for an extra systematic uncertainty related to changes in detector efficiency over time.

Presented in Fig.~\ref{fig:test} are the time-evolution of the backgrounds rates, the stability of the IBD-like excess (i.e.\@ events remaining after subtraction), and the daily pressure corrections applied. Good control over the background is maintained across the data-taking period and it is clear that fast neutrons constitute the largest background after selection. The significance of the IBD-like excess typically lies in the 3--5\textsigma{} range for reactor-on days and, with the exception of a few outliers, the reactor-off excess lies within 2\textsigma{} of zero. The variations in atmospheric pressure over this period result in adjustments of up to 75 events per day to the number of fast neutrons subtracted. The total number of antineutrino candidates is $\num{29479} \pm 603$ (stat.\@) with a signal-to-background ratio of 0.27.

%%%%%%%%%%%%%%%%%%%%%%%%%%%%
%%% OSCILLATION ANALYSIS %%%
%%%%%%%%%%%%%%%%%%%%%%%%%%%%

\begin{figure}
    \centering
    \includegraphics[width=1.0\linewidth]{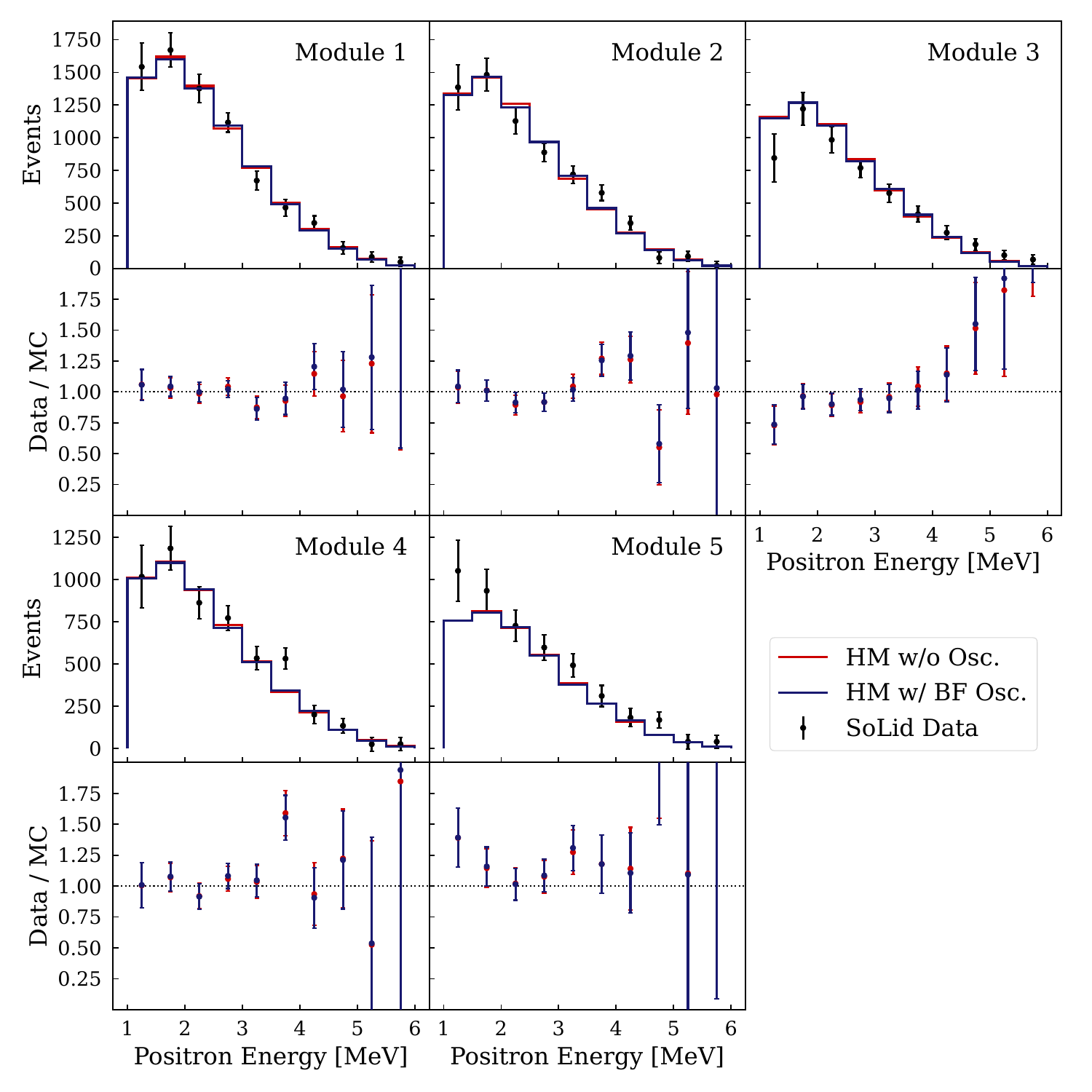}
    \caption{Reconstructed positron spectrum per detector module compared to the Huber-Mueller model without (red) and with (blue) the best-fit oscillations of the data applied. The two predictions have been scaled by their individually optimised normalisation parameter. Error bars represent combined statistical and systematic uncertainties.}
    \label{fig:spectra}
\end{figure}

A model-dependent search for oscillation of reactor antineutrinos has been conducted using frequentist and Bayesian methods. In both cases, the prediction for a particular oscillation scenario is built by re-weighting the MC events in the Huber-Mueller truth-space histogram, according to equation \ref{eq:survival}, and propagating the result to reconstructed space using a probability matrix \cite{remu2}. The reconstructed events are sorted into ten 500\,keV energy bins and five position bins.

The Bayesian analysis is based on a Markov chain Monte Carlo (MCMC) procedure that assumes a uniform prior for the oscillation amplitude \sthh{} (between 0.005 and 1) and a log-uniform prior for \dms{} (between 0.05 and 40\,\si{\electronvolt\squared}). Credible posterior regions are estimated with a Metropolis-Hasting sampling algorithm \cite{Metropolis1953,Hastings1970} based on a Poisson likelihood. The frequentist analysis follows the standard Feldman-Cousins technique \cite{Feldman1998}, with test statistic ${\Delta \chi^2 = \chi^2_{\text{H}_\text{x}} - \chi^2_{\text{H}_\text{BF}}}$ for oscillation hypothesis H$_\text{x}$. The best-fit value $\chi^2_{\text{H}_\text{BF}}$ is obtained by minimising globally over the sensitive region of the \dms{}-\sthh{} plane. We choose a $\chi^2$ metric of the form:

\begin{equation}\label{eq:chi}
   \chi^2 = r^{T} \: V_{\text{cov}}^{-1} \: r \:,
\end{equation}

\noindent where $V_{\text{cov}}$ is the matrix of statistical and systematic covariances and the residual $r$ is defined for position-energy bin $ij$ as:

\begin{equation}\label{eq:floatingnorm}
     r_{ij} = D_{ij} - \eta\: P_{ij} \:,
\end{equation}

\noindent where $D_{ij}$ and $P_{ij}$ refer to the data and prediction bin contents, respectively, and $\eta$ is a free-floating normalisation parameter which renders the fit insensitive to the overall antineutrino rate. Frequentist exclusion regions are thus defined by points in the discretised parameter-space grid whose $\Delta \chi^2$ exceeds some critical threshold at some confidence level (CL). The thresholds are calculated by generating $\mathcal{O}$(1000) pseudo-experiments at each grid point---via Cholesky decomposition of the covariance matrix---to build empirical distributions of the test statistic.

\begin{figure}[t]
    \centering
    \includegraphics[width=1.0\linewidth]{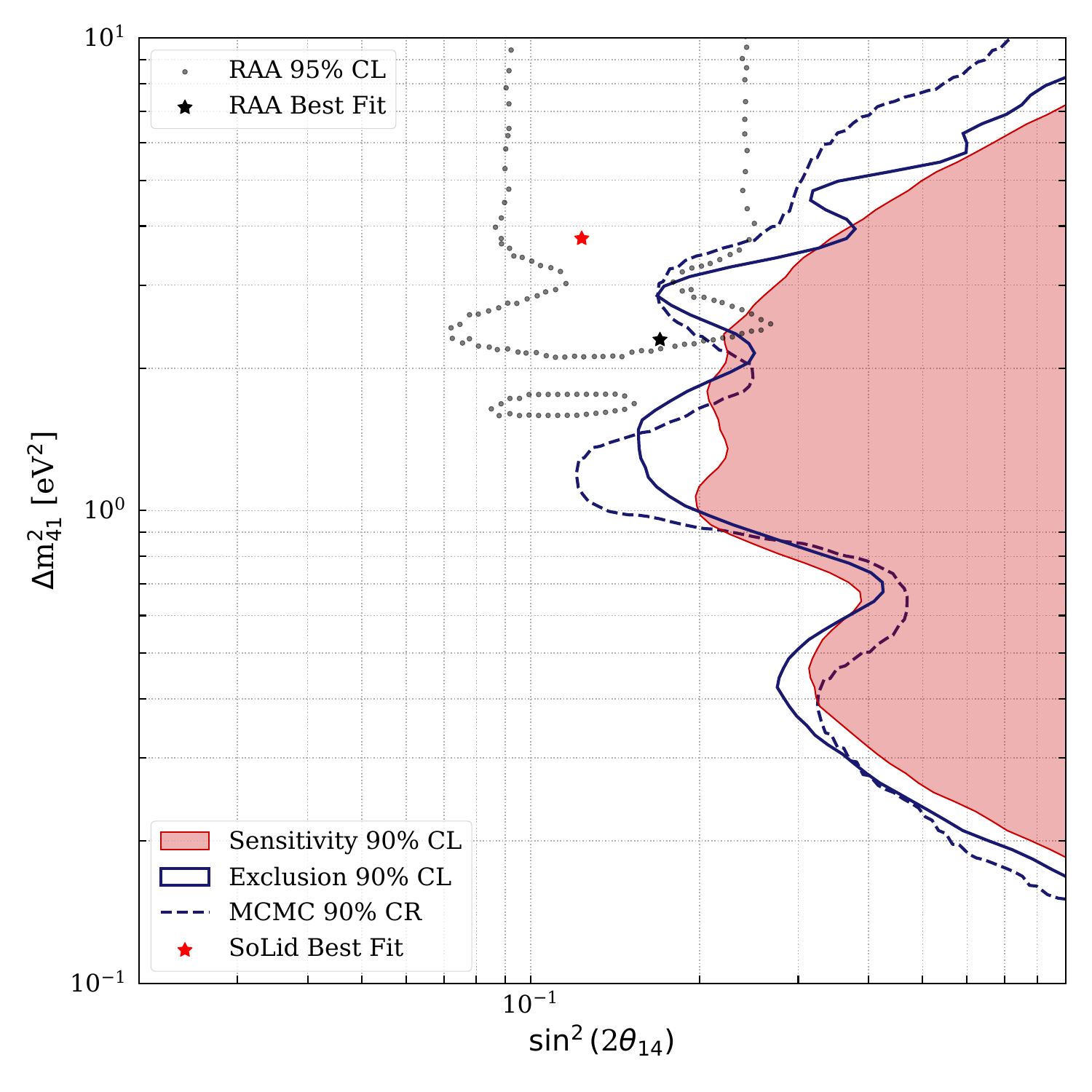}
    \caption{Frequentist exclusion and sensitivity contours (Feldman-Cousins) and Bayesian MCMC credible region. RAA contour and best-fit taken from Ref.~\cite{Mention2011}.}
    \label{fig:contour}
\end{figure}

The background-subtracted IBD spectra per module are shown in Fig.~\ref{fig:spectra}. The final spectrum follows an unblinding process in which the selection cuts were frozen prior to inspection of individual reactor cycles, which had to pass a data-MC goodness-of-fit check to be included. The prevailing systematic uncertainty originates from the precision of the PVT light yield---approximately 2\% on the first position-energy bin compared to 10\% statistical---but overall the bin errors are dominated by the statistical uncertainty. The reduced $\chi^2$ between the data and the no-oscillation HM prediction (H$_0$) is 49.8/49, indicating excellent agreement. Minimisation of the $\chi^2$ results in a $\Delta \chi^2$ of 0.915 at \sthh{}~=~0.123 and \dms{}~=~\SI{3.76}{\electronvolt\squared}, which cannot be rejected at 90\% CL. Therefore, the data exhibits a very mild preference for oscillations over the null hypothesis. Likewise, the Bayesian fit finds insufficient evidence to reject H$_0$ and identifies a most probable value at \sthh{}~=~0.02 and \dms{}~=~\SI{0.69}{\electronvolt\squared}. The frequentist exclusion contours and Bayesian credible regions (CR) are drawn in Fig.~\ref{fig:contour}. No significant evidence of an energy- and length-dependent oscillation pattern is observed and a portion of the 3+1 parameter space is rejected. The border of the credible region closely follows the shape of the frequentist contours but excludes a slightly greater area of parameter space. Our results are not sensitive to the Neutrino-4 best-fit point.

%%%%%%%%%%%%%%%%%%%%%%%%%%%%
%%% CONCLUSION & SUMMARY %%%
%%%%%%%%%%%%%%%%%%%%%%%%%%%%

In conclusion, this first analysis by the SoLid collaboration, conducted on a dataset comprising 280.3 live reactor-on days, has selected $\num{29479} \pm 603$ (stat.\@) antineutrino candidates with a signal-to-background ratio of 0.27. Uniquely amongst similarly motivated experiments, SoLid has direct access to the positron energy which permits the definition of high-level topological variables for event characterisation. Consequently, machine-learning tools for background suppression are both well-suited and have been necessary to achieve a reasonable background level given the initial rates. The stability of the detector response and background rates throughout the data-taking period were ensured by custom calibration tools. A model-dependent search for antineutrino disappearance has been performed, which finds no evidence of active-to-sterile oscillations in the 3+1 extension to the standard 3-neutrino paradigm and disfavours at 90\% CL a small section of the RAA allowed region.

This work was supported by the following funding sources: Science \& Technology Facilities Council (STFC), United Kingdom; FWO-Vlaanderen and the Vlaamse Herculesstichting, Belgium; Agence Nationale de la Recherche grant ANR-16CE31001803, Institut Carnot Mines, CNRS/IN2P3 and Region Pays de Loire, France. We are also deeply indebted to the SCK CEN staff for their generous assistance and accommodation. Individuals have received support from the FWO-Vlaanderen and the Belgian Federal Science Policy Office (BelSpo) under the IUAP network programme, the STFC Rutherford Fellowship program, and the European Research Council under the European Union’s Horizon 2020 Programme (H2020-CoG)/ERC Grant Agreement no.\@ 682474.

\bibliography{main}

\begin{thebibliography}{49}%
\makeatletter
\providecommand \@ifxundefined [1]{%
 \@ifx{#1\undefined}
}%
\providecommand \@ifnum [1]{%
 \ifnum #1\expandafter \@firstoftwo
 \else \expandafter \@secondoftwo
 \fi
}%
\providecommand \@ifx [1]{%
 \ifx #1\expandafter \@firstoftwo
 \else \expandafter \@secondoftwo
 \fi
}%
\providecommand \natexlab [1]{#1}%
\providecommand \enquote  [1]{``#1''}%
\providecommand \bibnamefont  [1]{#1}%
\providecommand \bibfnamefont [1]{#1}%
\providecommand \citenamefont [1]{#1}%
\providecommand \href@noop [0]{\@secondoftwo}%
\providecommand \href [0]{\begingroup \@sanitize@url \@href}%
\providecommand \@href[1]{\@@startlink{#1}\@@href}%
\providecommand \@@href[1]{\endgroup#1\@@endlink}%
\providecommand \@sanitize@url [0]{\catcode `\\12\catcode `\$12\catcode `\&12\catcode `\#12\catcode `\^12\catcode `\_12\catcode `\%12\relax}%
\providecommand \@@startlink[1]{}%
\providecommand \@@endlink[0]{}%
\providecommand \url  [0]{\begingroup\@sanitize@url \@url }%
\providecommand \@url [1]{\endgroup\@href {#1}{\urlprefix }}%
\providecommand \urlprefix  [0]{URL }%
\providecommand \Eprint [0]{\href }%
\providecommand \doibase [0]{https://doi.org/}%
\providecommand \selectlanguage [0]{\@gobble}%
\providecommand \bibinfo  [0]{\@secondoftwo}%
\providecommand \bibfield  [0]{\@secondoftwo}%
\providecommand \translation [1]{[#1]}%
\providecommand \BibitemOpen [0]{}%
\providecommand \bibitemStop [0]{}%
\providecommand \bibitemNoStop [0]{.\EOS\space}%
\providecommand \EOS [0]{\spacefactor3000\relax}%
\providecommand \BibitemShut  [1]{\csname bibitem#1\endcsname}%
\let\auto@bib@innerbib\@empty
%</preamble>
\bibitem [{\citenamefont {Mention}\ \emph {et~al.}(2011)\citenamefont {Mention}, \citenamefont {Fechner}, \citenamefont {Lasserre}, \citenamefont {Mueller}, \citenamefont {Lhuillier}, \citenamefont {Cribier},\ and\ \citenamefont {Letourneau}}]{Mention2011}%
  \BibitemOpen
  \bibfield  {author} {\bibinfo {author} {\bibfnamefont {G.}~\bibnamefont {Mention}}, \bibinfo {author} {\bibfnamefont {M.}~\bibnamefont {Fechner}}, \bibinfo {author} {\bibfnamefont {T.}~\bibnamefont {Lasserre}}, \bibinfo {author} {\bibfnamefont {T.~A.}\ \bibnamefont {Mueller}}, \bibinfo {author} {\bibfnamefont {D.}~\bibnamefont {Lhuillier}}, \bibinfo {author} {\bibfnamefont {M.}~\bibnamefont {Cribier}},\ and\ \bibinfo {author} {\bibfnamefont {A.}~\bibnamefont {Letourneau}},\ }\href {https://doi.org/10.1103/PhysRevD.83.073006} {\bibfield  {journal} {\bibinfo  {journal} {Physical Review D}\ }\textbf {\bibinfo {volume} {83}},\ \bibinfo {pages} {073006} (\bibinfo {year} {2011})}\BibitemShut {NoStop}%
\bibitem [{\citenamefont {Mueller}\ \emph {et~al.}(2011)\citenamefont {Mueller}, \citenamefont {Lhuillier}, \citenamefont {Fallot}, \citenamefont {Letourneau}, \citenamefont {Cormon}, \citenamefont {Fechner}, \citenamefont {Giot}, \citenamefont {Lasserre}, \citenamefont {Martino}, \citenamefont {Mention}, \citenamefont {Porta},\ and\ \citenamefont {Yermia}}]{Mueller2011}%
  \BibitemOpen
  \bibfield  {author} {\bibinfo {author} {\bibfnamefont {T.~A.}\ \bibnamefont {Mueller}}, \bibinfo {author} {\bibfnamefont {D.}~\bibnamefont {Lhuillier}}, \bibinfo {author} {\bibfnamefont {M.}~\bibnamefont {Fallot}}, \bibinfo {author} {\bibfnamefont {A.}~\bibnamefont {Letourneau}}, \bibinfo {author} {\bibfnamefont {S.}~\bibnamefont {Cormon}}, \bibinfo {author} {\bibfnamefont {M.}~\bibnamefont {Fechner}}, \bibinfo {author} {\bibfnamefont {L.}~\bibnamefont {Giot}}, \bibinfo {author} {\bibfnamefont {T.}~\bibnamefont {Lasserre}}, \bibinfo {author} {\bibfnamefont {J.}~\bibnamefont {Martino}}, \bibinfo {author} {\bibfnamefont {G.}~\bibnamefont {Mention}}, \bibinfo {author} {\bibfnamefont {A.}~\bibnamefont {Porta}},\ and\ \bibinfo {author} {\bibfnamefont {F.}~\bibnamefont {Yermia}},\ }\href {https://doi.org/10.1103/PhysRevC.83.054615} {\bibfield  {journal} {\bibinfo  {journal} {Physical Review C}\ }\textbf {\bibinfo {volume} {83}},\ \bibinfo {pages} {054615} (\bibinfo {year} {2011})}\BibitemShut {NoStop}%
\bibitem [{\citenamefont {Huber}(2011)}]{Huber2011}%
  \BibitemOpen
  \bibfield  {author} {\bibinfo {author} {\bibfnamefont {P.}~\bibnamefont {Huber}},\ }\href {https://doi.org/10.1103/PhysRevC.84.024617} {\bibfield  {journal} {\bibinfo  {journal} {Physical Review C}\ }\textbf {\bibinfo {volume} {84}},\ \bibinfo {pages} {024617} (\bibinfo {year} {2011})}\BibitemShut {NoStop}%
\bibitem [{\citenamefont {Hayen}\ \emph {et~al.}(2019)\citenamefont {Hayen}, \citenamefont {Kostensalo}, \citenamefont {Severijns},\ and\ \citenamefont {Suhonen}}]{Hayen2019}%
  \BibitemOpen
  \bibfield  {author} {\bibinfo {author} {\bibfnamefont {L.}~\bibnamefont {Hayen}}, \bibinfo {author} {\bibfnamefont {J.}~\bibnamefont {Kostensalo}}, \bibinfo {author} {\bibfnamefont {N.}~\bibnamefont {Severijns}},\ and\ \bibinfo {author} {\bibfnamefont {J.}~\bibnamefont {Suhonen}},\ }\href {https://doi.org/10.1103/PhysRevC.99.031301} {\bibfield  {journal} {\bibinfo  {journal} {Physical Review C}\ }\textbf {\bibinfo {volume} {99}},\ \bibinfo {pages} {031301} (\bibinfo {year} {2019})}\BibitemShut {NoStop}%
\bibitem [{\citenamefont {Li}\ and\ \citenamefont {Zhang}(2019)}]{Li2019}%
  \BibitemOpen
  \bibfield  {author} {\bibinfo {author} {\bibfnamefont {Y.-F.}\ \bibnamefont {Li}}\ and\ \bibinfo {author} {\bibfnamefont {D.}~\bibnamefont {Zhang}},\ }\href {https://doi.org/10.1103/PhysRevD.100.053005} {\bibfield  {journal} {\bibinfo  {journal} {Physical Review D}\ }\textbf {\bibinfo {volume} {100}},\ \bibinfo {pages} {053005} (\bibinfo {year} {2019})}\BibitemShut {NoStop}%
\bibitem [{\citenamefont {Kopeikin}\ \emph {et~al.}(2021)\citenamefont {Kopeikin}, \citenamefont {Skorokhvatov},\ and\ \citenamefont {Titov}}]{Kopeikin2021}%
  \BibitemOpen
  \bibfield  {author} {\bibinfo {author} {\bibfnamefont {V.}~\bibnamefont {Kopeikin}}, \bibinfo {author} {\bibfnamefont {M.}~\bibnamefont {Skorokhvatov}},\ and\ \bibinfo {author} {\bibfnamefont {O.}~\bibnamefont {Titov}},\ }\href {https://doi.org/10.1103/PhysRevD.104.L071301} {\bibfield  {journal} {\bibinfo  {journal} {Physical Review D}\ }\textbf {\bibinfo {volume} {104}},\ \bibinfo {pages} {L071301} (\bibinfo {year} {2021})}\BibitemShut {NoStop}%
\bibitem [{\citenamefont {Estienne}\ \emph {et~al.}(2019)\citenamefont {Estienne} \emph {et~al.}}]{Estienne2019}%
  \BibitemOpen
  \bibfield  {author} {\bibinfo {author} {\bibfnamefont {M.}~\bibnamefont {Estienne}} \emph {et~al.},\ }\href {https://doi.org/10.1103/PhysRevLett.123.022502} {\bibfield  {journal} {\bibinfo  {journal} {Physical Review Letters}\ }\textbf {\bibinfo {volume} {123}},\ \bibinfo {pages} {022502} (\bibinfo {year} {2019})}\BibitemShut {NoStop}%
\bibitem [{\citenamefont {Perissé}\ \emph {et~al.}(2023)\citenamefont {Perissé}, \citenamefont {Onillon}, \citenamefont {Mougeot}, \citenamefont {Vivier}, \citenamefont {Lasserre}, \citenamefont {Letourneau}, \citenamefont {Lhuillier},\ and\ \citenamefont {Mention}}]{Perisse2023}%
  \BibitemOpen
  \bibfield  {author} {\bibinfo {author} {\bibfnamefont {L.}~\bibnamefont {Perissé}}, \bibinfo {author} {\bibfnamefont {A.}~\bibnamefont {Onillon}}, \bibinfo {author} {\bibfnamefont {X.}~\bibnamefont {Mougeot}}, \bibinfo {author} {\bibfnamefont {M.}~\bibnamefont {Vivier}}, \bibinfo {author} {\bibfnamefont {T.}~\bibnamefont {Lasserre}}, \bibinfo {author} {\bibfnamefont {A.}~\bibnamefont {Letourneau}}, \bibinfo {author} {\bibfnamefont {D.}~\bibnamefont {Lhuillier}},\ and\ \bibinfo {author} {\bibfnamefont {G.}~\bibnamefont {Mention}},\ }\href {https://doi.org/10.1103/PhysRevC.108.055501} {\bibfield  {journal} {\bibinfo  {journal} {Phys. Rev. C}\ }\textbf {\bibinfo {volume} {108}},\ \bibinfo {pages} {055501} (\bibinfo {year} {2023})}\BibitemShut {NoStop}%
\bibitem [{\citenamefont {Letourneau}\ \emph {et~al.}(2023)\citenamefont {Letourneau}, \citenamefont {Savu}, \citenamefont {Lhuillier}, \citenamefont {Lasserre}, \citenamefont {Materna}, \citenamefont {Mention}, \citenamefont {Mougeot}, \citenamefont {Onillon}, \citenamefont {Perisse},\ and\ \citenamefont {Vivier}}]{Letourneau2023}%
  \BibitemOpen
  \bibfield  {author} {\bibinfo {author} {\bibfnamefont {A.}~\bibnamefont {Letourneau}}, \bibinfo {author} {\bibfnamefont {V.}~\bibnamefont {Savu}}, \bibinfo {author} {\bibfnamefont {D.}~\bibnamefont {Lhuillier}}, \bibinfo {author} {\bibfnamefont {T.}~\bibnamefont {Lasserre}}, \bibinfo {author} {\bibfnamefont {T.}~\bibnamefont {Materna}}, \bibinfo {author} {\bibfnamefont {G.}~\bibnamefont {Mention}}, \bibinfo {author} {\bibfnamefont {X.}~\bibnamefont {Mougeot}}, \bibinfo {author} {\bibfnamefont {A.}~\bibnamefont {Onillon}}, \bibinfo {author} {\bibfnamefont {L.}~\bibnamefont {Perisse}},\ and\ \bibinfo {author} {\bibfnamefont {M.}~\bibnamefont {Vivier}},\ }\href {https://doi.org/10.1103/PhysRevLett.130.021801} {\bibfield  {journal} {\bibinfo  {journal} {Physical Review Letters}\ }\textbf {\bibinfo {volume} {130}},\ \bibinfo {pages} {021801} (\bibinfo {year} {2023})}\BibitemShut {NoStop}%
\bibitem [{\citenamefont {Giunti}\ \emph {et~al.}(2022)\citenamefont {Giunti}, \citenamefont {Li}, \citenamefont {Ternes},\ and\ \citenamefont {Xin}}]{Giunti2022}%
  \BibitemOpen
  \bibfield  {author} {\bibinfo {author} {\bibfnamefont {C.}~\bibnamefont {Giunti}}, \bibinfo {author} {\bibfnamefont {Y.}~\bibnamefont {Li}}, \bibinfo {author} {\bibfnamefont {C.}~\bibnamefont {Ternes}},\ and\ \bibinfo {author} {\bibfnamefont {Z.}~\bibnamefont {Xin}},\ }\href {https://doi.org/10.1016/j.physletb.2022.137054} {\bibfield  {journal} {\bibinfo  {journal} {Physics Letters B}\ }\textbf {\bibinfo {volume} {829}},\ \bibinfo {pages} {137054} (\bibinfo {year} {2022})}\BibitemShut {NoStop}%
\bibitem [{\citenamefont {Schreckenbach}\ \emph {et~al.}(1981)\citenamefont {Schreckenbach}, \citenamefont {Faust}, \citenamefont {von Feilitzsch}, \citenamefont {Hahn}, \citenamefont {Hawerkamp},\ and\ \citenamefont {Vuilleumier}}]{Schreckenbach1981}%
  \BibitemOpen
  \bibfield  {author} {\bibinfo {author} {\bibfnamefont {K.}~\bibnamefont {Schreckenbach}}, \bibinfo {author} {\bibfnamefont {H.}~\bibnamefont {Faust}}, \bibinfo {author} {\bibfnamefont {F.}~\bibnamefont {von Feilitzsch}}, \bibinfo {author} {\bibfnamefont {A.}~\bibnamefont {Hahn}}, \bibinfo {author} {\bibfnamefont {K.}~\bibnamefont {Hawerkamp}},\ and\ \bibinfo {author} {\bibfnamefont {J.}~\bibnamefont {Vuilleumier}},\ }\href {https://doi.org/10.1016/0370-2693(81)91120-5} {\bibfield  {journal} {\bibinfo  {journal} {Physics Letters B}\ }\textbf {\bibinfo {volume} {99}},\ \bibinfo {pages} {251} (\bibinfo {year} {1981})}\BibitemShut {NoStop}%
\bibitem [{\citenamefont {von Feilitzsch}\ \emph {et~al.}(1982)\citenamefont {von Feilitzsch}, \citenamefont {Hahn},\ and\ \citenamefont {Schreckenbach}}]{Feilitzsch1982}%
  \BibitemOpen
  \bibfield  {author} {\bibinfo {author} {\bibfnamefont {F.}~\bibnamefont {von Feilitzsch}}, \bibinfo {author} {\bibfnamefont {A.}~\bibnamefont {Hahn}},\ and\ \bibinfo {author} {\bibfnamefont {K.}~\bibnamefont {Schreckenbach}},\ }\href {https://doi.org/10.1016/0370-2693(82)90622-0} {\bibfield  {journal} {\bibinfo  {journal} {Physics Letters B}\ }\textbf {\bibinfo {volume} {118}},\ \bibinfo {pages} {162} (\bibinfo {year} {1982})}\BibitemShut {NoStop}%
\bibitem [{\citenamefont {Schreckenbach}\ \emph {et~al.}(1985)\citenamefont {Schreckenbach}, \citenamefont {Colvin}, \citenamefont {Gelletly},\ and\ \citenamefont {Feilitzsch}}]{Schreckenbach1985}%
  \BibitemOpen
  \bibfield  {author} {\bibinfo {author} {\bibfnamefont {K.}~\bibnamefont {Schreckenbach}}, \bibinfo {author} {\bibfnamefont {G.}~\bibnamefont {Colvin}}, \bibinfo {author} {\bibfnamefont {W.}~\bibnamefont {Gelletly}},\ and\ \bibinfo {author} {\bibfnamefont {F.~V.}\ \bibnamefont {Feilitzsch}},\ }\href {https://doi.org/10.1016/0370-2693(85)91337-1} {\bibfield  {journal} {\bibinfo  {journal} {Physics Letters B}\ }\textbf {\bibinfo {volume} {160}},\ \bibinfo {pages} {325} (\bibinfo {year} {1985})}\BibitemShut {NoStop}%
\bibitem [{\citenamefont {Hayes}\ and\ \citenamefont {Vogel}(2016)}]{Hayes2016}%
  \BibitemOpen
  \bibfield  {author} {\bibinfo {author} {\bibfnamefont {A.~C.}\ \bibnamefont {Hayes}}\ and\ \bibinfo {author} {\bibfnamefont {P.}~\bibnamefont {Vogel}},\ }\href {https://doi.org/10.1146/annurev-nucl-102115-044826} {\bibfield  {journal} {\bibinfo  {journal} {Annual Review of Nuclear and Particle Science}\ }\textbf {\bibinfo {volume} {66}},\ \bibinfo {pages} {219} (\bibinfo {year} {2016})}\BibitemShut {NoStop}%
\bibitem [{\citenamefont {Sonzogni}\ \emph {et~al.}(2016)\citenamefont {Sonzogni}, \citenamefont {McCutchan}, \citenamefont {Johnson},\ and\ \citenamefont {Dimitriou}}]{Sonzogni2016}%
  \BibitemOpen
  \bibfield  {author} {\bibinfo {author} {\bibfnamefont {A.~A.}\ \bibnamefont {Sonzogni}}, \bibinfo {author} {\bibfnamefont {E.~A.}\ \bibnamefont {McCutchan}}, \bibinfo {author} {\bibfnamefont {T.~D.}\ \bibnamefont {Johnson}},\ and\ \bibinfo {author} {\bibfnamefont {P.}~\bibnamefont {Dimitriou}},\ }\href {https://doi.org/10.1103/PhysRevLett.116.132502} {\bibfield  {journal} {\bibinfo  {journal} {Physical Review Letters}\ }\textbf {\bibinfo {volume} {116}},\ \bibinfo {pages} {132502} (\bibinfo {year} {2016})}\BibitemShut {NoStop}%
\bibitem [{\citenamefont {An}\ \emph {et~al.}(2017)\citenamefont {An} \emph {et~al.}}]{An2017}%
  \BibitemOpen
  \bibfield  {author} {\bibinfo {author} {\bibfnamefont {F.~P.}\ \bibnamefont {An}} \emph {et~al.},\ }\href {https://doi.org/10.1103/PhysRevLett.118.251801} {\bibfield  {journal} {\bibinfo  {journal} {Physical Review Letters}\ }\textbf {\bibinfo {volume} {118}},\ \bibinfo {pages} {251801} (\bibinfo {year} {2017})}\BibitemShut {NoStop}%
\bibitem [{\citenamefont {An}\ \emph {et~al.}(2023)\citenamefont {An} \emph {et~al.}}]{An2023}%
  \BibitemOpen
  \bibfield  {author} {\bibinfo {author} {\bibfnamefont {F.~P.}\ \bibnamefont {An}} \emph {et~al.},\ }\href {https://doi.org/10.1103/PhysRevLett.130.211801} {\bibfield  {journal} {\bibinfo  {journal} {Physical Review Letters}\ }\textbf {\bibinfo {volume} {130}},\ \bibinfo {pages} {211801} (\bibinfo {year} {2023})}\BibitemShut {NoStop}%
\bibitem [{\citenamefont {Bak}\ \emph {et~al.}(2019)\citenamefont {Bak} \emph {et~al.}}]{Bak2019}%
  \BibitemOpen
  \bibfield  {author} {\bibinfo {author} {\bibfnamefont {G.}~\bibnamefont {Bak}} \emph {et~al.},\ }\href {https://doi.org/10.1103/PhysRevLett.122.232501} {\bibfield  {journal} {\bibinfo  {journal} {Physical Review Letters}\ }\textbf {\bibinfo {volume} {122}},\ \bibinfo {pages} {232501} (\bibinfo {year} {2019})}\BibitemShut {NoStop}%
\bibitem [{\citenamefont {Andriamirado}\ \emph {et~al.}(2021)\citenamefont {Andriamirado} \emph {et~al.}}]{prospect2}%
  \BibitemOpen
  \bibfield  {author} {\bibinfo {author} {\bibfnamefont {M.}~\bibnamefont {Andriamirado}} \emph {et~al.},\ }\href {https://doi.org/10.1103/PhysRevD.103.032001} {\bibfield  {journal} {\bibinfo  {journal} {Physical Review D}\ }\textbf {\bibinfo {volume} {103}},\ \bibinfo {pages} {032001} (\bibinfo {year} {2021})}\BibitemShut {NoStop}%
\bibitem [{\citenamefont {Almazán}\ \emph {et~al.}(2020{\natexlab{a}})\citenamefont {Almazán} \emph {et~al.}}]{stereo2}%
  \BibitemOpen
  \bibfield  {author} {\bibinfo {author} {\bibfnamefont {H.}~\bibnamefont {Almazán}} \emph {et~al.},\ }\href {https://doi.org/10.1103/PhysRevD.102.052002} {\bibfield  {journal} {\bibinfo  {journal} {Physical Review D}\ }\textbf {\bibinfo {volume} {102}},\ \bibinfo {pages} {052002} (\bibinfo {year} {2020}{\natexlab{a}})}\BibitemShut {NoStop}%
\bibitem [{\citenamefont {Serebrov}\ \emph {et~al.}(2017)\citenamefont {Serebrov} \emph {et~al.}}]{Serebrov2017}%
  \BibitemOpen
  \bibfield  {author} {\bibinfo {author} {\bibfnamefont {A.~P.}\ \bibnamefont {Serebrov}} \emph {et~al.},\ }\href {https://doi.org/10.1088/1742-6596/888/1/012089} {\bibfield  {journal} {\bibinfo  {journal} {Journal of Physics: Conference Series}\ }\textbf {\bibinfo {volume} {888}},\ \bibinfo {pages} {012089} (\bibinfo {year} {2017})}\BibitemShut {NoStop}%
\bibitem [{\citenamefont {Alekseev}\ \emph {et~al.}(2018)\citenamefont {Alekseev} \emph {et~al.}}]{Alekseev2018}%
  \BibitemOpen
  \bibfield  {author} {\bibinfo {author} {\bibfnamefont {I.}~\bibnamefont {Alekseev}} \emph {et~al.},\ }\href {https://doi.org/10.1016/j.physletb.2018.10.038} {\bibfield  {journal} {\bibinfo  {journal} {Physics Letters B}\ }\textbf {\bibinfo {volume} {787}},\ \bibinfo {pages} {56} (\bibinfo {year} {2018})}\BibitemShut {NoStop}%
\bibitem [{\citenamefont {Ko}\ \emph {et~al.}(2017)\citenamefont {Ko} \emph {et~al.}}]{Ko2017}%
  \BibitemOpen
  \bibfield  {author} {\bibinfo {author} {\bibfnamefont {Y.~J.}\ \bibnamefont {Ko}} \emph {et~al.},\ }\href {https://doi.org/10.1103/PhysRevLett.118.121802} {\bibfield  {journal} {\bibinfo  {journal} {Physical Review Letters}\ }\textbf {\bibinfo {volume} {118}},\ \bibinfo {pages} {121802} (\bibinfo {year} {2017})}\BibitemShut {NoStop}%
\bibitem [{\citenamefont {Serebrov}\ \emph {et~al.}(2021)\citenamefont {Serebrov} \emph {et~al.}}]{Serebrov2021}%
  \BibitemOpen
  \bibfield  {author} {\bibinfo {author} {\bibfnamefont {A.~P.}\ \bibnamefont {Serebrov}} \emph {et~al.},\ }\href {https://doi.org/10.1103/PhysRevD.104.032003} {\bibfield  {journal} {\bibinfo  {journal} {Physical Review D}\ }\textbf {\bibinfo {volume} {104}},\ \bibinfo {pages} {032003} (\bibinfo {year} {2021})}\BibitemShut {NoStop}%
\bibitem [{\citenamefont {Almazán}\ \emph {et~al.}(2020{\natexlab{b}})\citenamefont {Almazán} \emph {et~al.}}]{badnu4a}%
  \BibitemOpen
  \bibfield  {author} {\bibinfo {author} {\bibfnamefont {H.}~\bibnamefont {Almazán}} \emph {et~al.},\ }\bibfield  {journal} {\bibinfo  {journal} {arXiv e-prints}\ }\href {https://doi.org/10.48550/arXiv.2006.13147} {10.48550/arXiv.2006.13147} (\bibinfo {year} {2020}{\natexlab{b}})\BibitemShut {NoStop}%
\bibitem [{\citenamefont {Abazov}\ \emph {et~al.}(1991)\citenamefont {Abazov} \emph {et~al.}}]{Abazov1991}%
  \BibitemOpen
  \bibfield  {author} {\bibinfo {author} {\bibfnamefont {A.~I.}\ \bibnamefont {Abazov}} \emph {et~al.},\ }\href {https://doi.org/10.1103/PhysRevLett.67.3332} {\bibfield  {journal} {\bibinfo  {journal} {Physical Review Letters}\ }\textbf {\bibinfo {volume} {67}},\ \bibinfo {pages} {3332} (\bibinfo {year} {1991})}\BibitemShut {NoStop}%
\bibitem [{\citenamefont {Hampel}\ \emph {et~al.}(1998)\citenamefont {Hampel}, \citenamefont {Heusser}, \citenamefont {Kiko}, \citenamefont {Kirsten}, \citenamefont {Laubenstein}, \citenamefont {Pernicka}, \citenamefont {Rau}, \citenamefont {Rönn}, \citenamefont {Schlosser}, \citenamefont {Wójcik}, \citenamefont {v.~Ammon}, \citenamefont {Ebert}, \citenamefont {Fritsch}, \citenamefont {Heidt}, \citenamefont {Henrich} \emph {et~al.}}]{Hampel1998}%
  \BibitemOpen
  \bibfield  {author} {\bibinfo {author} {\bibfnamefont {W.}~\bibnamefont {Hampel}}, \bibinfo {author} {\bibfnamefont {G.}~\bibnamefont {Heusser}}, \bibinfo {author} {\bibfnamefont {J.}~\bibnamefont {Kiko}}, \bibinfo {author} {\bibfnamefont {T.}~\bibnamefont {Kirsten}}, \bibinfo {author} {\bibfnamefont {M.}~\bibnamefont {Laubenstein}}, \bibinfo {author} {\bibfnamefont {E.}~\bibnamefont {Pernicka}}, \bibinfo {author} {\bibfnamefont {W.}~\bibnamefont {Rau}}, \bibinfo {author} {\bibfnamefont {U.}~\bibnamefont {Rönn}}, \bibinfo {author} {\bibfnamefont {C.}~\bibnamefont {Schlosser}}, \bibinfo {author} {\bibfnamefont {M.}~\bibnamefont {Wójcik}}, \bibinfo {author} {\bibfnamefont {R.}~\bibnamefont {v.~Ammon}}, \bibinfo {author} {\bibfnamefont {K.}~\bibnamefont {Ebert}}, \bibinfo {author} {\bibfnamefont {T.}~\bibnamefont {Fritsch}}, \bibinfo {author} {\bibfnamefont {D.}~\bibnamefont {Heidt}}, \bibinfo {author} {\bibfnamefont {E.}~\bibnamefont {Henrich}}, \emph {et~al.},\ }\href
  {https://doi.org/10.1016/S0370-2693(97)01562-1} {\bibfield  {journal} {\bibinfo  {journal} {Physics Letters B}\ }\textbf {\bibinfo {volume} {420}},\ \bibinfo {pages} {114} (\bibinfo {year} {1998})}\BibitemShut {NoStop}%
\bibitem [{\citenamefont {Barinov}\ \emph {et~al.}(2022)\citenamefont {Barinov} \emph {et~al.}}]{Barinov2022}%
  \BibitemOpen
  \bibfield  {author} {\bibinfo {author} {\bibfnamefont {V.~V.}\ \bibnamefont {Barinov}} \emph {et~al.},\ }\href {https://doi.org/10.1103/PhysRevC.105.065502} {\bibfield  {journal} {\bibinfo  {journal} {Physical Review C}\ }\textbf {\bibinfo {volume} {105}},\ \bibinfo {pages} {065502} (\bibinfo {year} {2022})}\BibitemShut {NoStop}%
\bibitem [{\citenamefont {Abreu}\ \emph {et~al.}(2017)\citenamefont {Abreu} \emph {et~al.}}]{Abreu2017}%
  \BibitemOpen
  \bibfield  {author} {\bibinfo {author} {\bibfnamefont {Y.}~\bibnamefont {Abreu}} \emph {et~al.},\ }\href {https://doi.org/10.1088/1748-0221/12/04/P04024} {\bibfield  {journal} {\bibinfo  {journal} {Journal of Instrumentation}\ }\textbf {\bibinfo {volume} {12}}\bibinfo  {number} { (04)},\ \bibinfo {pages} {P04024}}\BibitemShut {NoStop}%
\bibitem [{\citenamefont {Abreu}\ \emph {et~al.}(2018{\natexlab{a}})\citenamefont {Abreu} \emph {et~al.}}]{Abreu2018}%
  \BibitemOpen
\bibfield  {number} {  }\bibfield  {author} {\bibinfo {author} {\bibfnamefont {Y.}~\bibnamefont {Abreu}} \emph {et~al.},\ }\href {https://doi.org/10.1088/1748-0221/13/05/P05005} {\bibfield  {journal} {\bibinfo  {journal} {Journal of Instrumentation}\ }\textbf {\bibinfo {volume} {13}}\bibinfo  {number} { (05)},\ \bibinfo {pages} {P05005}}\BibitemShut {NoStop}%
\bibitem [{\citenamefont {Abreu}\ \emph {et~al.}(2018{\natexlab{b}})\citenamefont {Abreu} \emph {et~al.}}]{Abreu2018v2}%
  \BibitemOpen
\bibfield  {number} {  }\bibfield  {author} {\bibinfo {author} {\bibfnamefont {Y.}~\bibnamefont {Abreu}} \emph {et~al.},\ }\href {https://doi.org/10.1088/1748-0221/13/09/P09005} {\bibfield  {journal} {\bibinfo  {journal} {Journal of Instrumentation}\ }\textbf {\bibinfo {volume} {13}}\bibinfo  {number} { (09)},\ \bibinfo {pages} {P09005}}\BibitemShut {NoStop}%
\bibitem [{\citenamefont {Abreu}\ \emph {et~al.}(2019{\natexlab{a}})\citenamefont {Abreu} \emph {et~al.}}]{Abreu2019v2}%
  \BibitemOpen
\bibfield  {number} {  }\bibfield  {author} {\bibinfo {author} {\bibfnamefont {Y.}~\bibnamefont {Abreu}} \emph {et~al.},\ }\href {https://doi.org/10.1088/1748-0221/14/02/P02014} {\bibfield  {journal} {\bibinfo  {journal} {Journal of Instrumentation}\ }\textbf {\bibinfo {volume} {14}}\bibinfo  {number} { (02)},\ \bibinfo {pages} {P02014}}\BibitemShut {NoStop}%
\bibitem [{\citenamefont {Abreu}\ \emph {et~al.}(2019{\natexlab{b}})\citenamefont {Abreu} \emph {et~al.}}]{Abreu2019}%
  \BibitemOpen
\bibfield  {number} {  }\bibfield  {author} {\bibinfo {author} {\bibfnamefont {Y.}~\bibnamefont {Abreu}} \emph {et~al.},\ }\href {https://doi.org/10.1088/1748-0221/14/11/P11003} {\bibfield  {journal} {\bibinfo  {journal} {Journal of Instrumentation}\ }\textbf {\bibinfo {volume} {14}}\bibinfo  {number} { (11)},\ \bibinfo {pages} {P11003}}\BibitemShut {NoStop}%
\bibitem [{\citenamefont {Abreu}\ \emph {et~al.}(2021)\citenamefont {Abreu} \emph {et~al.}}]{Abreu2021}%
  \BibitemOpen
\bibfield  {number} {  }\bibfield  {author} {\bibinfo {author} {\bibfnamefont {Y.}~\bibnamefont {Abreu}} \emph {et~al.},\ }\href {https://doi.org/10.1088/1748-0221/16/02/P02025} {\bibfield  {journal} {\bibinfo  {journal} {Journal of Instrumentation}\ }\textbf {\bibinfo {volume} {16}}\bibinfo  {number} { (02)},\ \bibinfo {pages} {P02025}}\BibitemShut {NoStop}%
\bibitem [{\citenamefont {Reines}\ and\ \citenamefont {Cowan}(1953)}]{Reines1953v2}%
  \BibitemOpen
\bibfield  {number} {  }\bibfield  {author} {\bibinfo {author} {\bibfnamefont {F.}~\bibnamefont {Reines}}\ and\ \bibinfo {author} {\bibfnamefont {C.~L.}\ \bibnamefont {Cowan}},\ }\bibfield  {journal} {\bibinfo  {journal} {Physical Review}\ }\textbf {\bibinfo {volume} {92}},\ \href {https://doi.org/10.1103/PhysRev.92.830} {10.1103/PhysRev.92.830} (\bibinfo {year} {1953})\BibitemShut {NoStop}%
\bibitem [{\citenamefont {Simón}\ \emph {et~al.}(2016)\citenamefont {Simón}, \citenamefont {Ferrario},\ and\ \citenamefont {Izmaylov}}]{next}%
  \BibitemOpen
  \bibfield  {author} {\bibinfo {author} {\bibfnamefont {A.}~\bibnamefont {Simón}}, \bibinfo {author} {\bibfnamefont {P.}~\bibnamefont {Ferrario}},\ and\ \bibinfo {author} {\bibfnamefont {A.}~\bibnamefont {Izmaylov}},\ }\href {https://doi.org/10.1016/j.nuclphysbps.2015.10.010} {\bibfield  {journal} {\bibinfo  {journal} {Nuclear and Particle Physics Proceedings}\ }\textbf {\bibinfo {volume} {273-275}},\ \bibinfo {pages} {2624} (\bibinfo {year} {2016})}\BibitemShut {NoStop}%
\bibitem [{\citenamefont {Mallat}\ and\ \citenamefont {Zhang}(1993)}]{omp}%
  \BibitemOpen
  \bibfield  {author} {\bibinfo {author} {\bibfnamefont {S.}~\bibnamefont {Mallat}}\ and\ \bibinfo {author} {\bibfnamefont {Z.}~\bibnamefont {Zhang}},\ }\href {https://doi.org/10.1109/78.258082} {\bibfield  {journal} {\bibinfo  {journal} {IEEE Transactions on Signal Processing}\ }\textbf {\bibinfo {volume} {41}},\ \bibinfo {pages} {3397} (\bibinfo {year} {1993})}\BibitemShut {NoStop}%
\bibitem [{\citenamefont {Abreu}\ \emph {et~al.}(2024)\citenamefont {Abreu} \emph {et~al.}}]{Abreu2024}%
  \BibitemOpen
  \bibfield  {author} {\bibinfo {author} {\bibfnamefont {Y.}~\bibnamefont {Abreu}} \emph {et~al.},\ }\bibfield  {journal} {\bibinfo  {journal} {arXiv e-prints}\ }\href {https://doi.org/https://doi.org/10.48550/arXiv.2404.03580} {https://doi.org/10.48550/arXiv.2404.03580} (\bibinfo {year} {2024})\BibitemShut {NoStop}%
\bibitem [{\citenamefont {Yeresko}(2022)}]{mike}%
  \BibitemOpen
  \bibfield  {author} {\bibinfo {author} {\bibfnamefont {M.}~\bibnamefont {Yeresko}},\ }\href {https://theses.hal.science/tel-04433698} {Ph.D. thesis},\ \bibinfo  {school} {{Universit{\'e} Clermont Auvergne}} (\bibinfo {year} {2022})\BibitemShut {NoStop}%
\bibitem [{\citenamefont {Agostinelli}\ \emph {et~al.}(2003)\citenamefont {Agostinelli} \emph {et~al.}}]{geant4}%
  \BibitemOpen
  \bibfield  {author} {\bibinfo {author} {\bibfnamefont {S.}~\bibnamefont {Agostinelli}} \emph {et~al.},\ }\href {https://doi.org/10.1016/S0168-9002(03)01368-8} {\bibfield  {journal} {\bibinfo  {journal} {Nuclear Instruments and Methods in Physics Research Section A: Accelerators, Spectrometers, Detectors and Associated Equipment}\ }\textbf {\bibinfo {volume} {506}},\ \bibinfo {pages} {250} (\bibinfo {year} {2003})}\BibitemShut {NoStop}%
\bibitem [{\citenamefont {Waters}\ \emph {et~al.}(2007)\citenamefont {Waters}, \citenamefont {McKinney}, \citenamefont {Durkee}, \citenamefont {Fensin}, \citenamefont {Hendricks}, \citenamefont {James}, \citenamefont {Johns},\ and\ \citenamefont {Pelowitz}}]{mcnp}%
  \BibitemOpen
  \bibfield  {author} {\bibinfo {author} {\bibfnamefont {L.~S.}\ \bibnamefont {Waters}}, \bibinfo {author} {\bibfnamefont {G.~W.}\ \bibnamefont {McKinney}}, \bibinfo {author} {\bibfnamefont {J.~W.}\ \bibnamefont {Durkee}}, \bibinfo {author} {\bibfnamefont {M.~L.}\ \bibnamefont {Fensin}}, \bibinfo {author} {\bibfnamefont {J.~S.}\ \bibnamefont {Hendricks}}, \bibinfo {author} {\bibfnamefont {M.~R.}\ \bibnamefont {James}}, \bibinfo {author} {\bibfnamefont {R.~C.}\ \bibnamefont {Johns}},\ and\ \bibinfo {author} {\bibfnamefont {D.~B.}\ \bibnamefont {Pelowitz}}\ }(\bibinfo  {publisher} {AIP},\ \bibinfo {year} {2007})\ pp.\ \bibinfo {pages} {81--90}\BibitemShut {NoStop}%
\bibitem [{\citenamefont {Méplan}\ \emph {et~al.}(2009)\citenamefont {Méplan} \emph {et~al.}}]{mure}%
  \BibitemOpen
  \bibfield  {author} {\bibinfo {author} {\bibfnamefont {O.}~\bibnamefont {Méplan}} \emph {et~al.},\ }\href {http://www.nea.fr/tools/abstract/detail/nea-1845} {\bibfield  {journal} {\bibinfo  {journal} {Tech. Rep., LPSC 0912 and IPNO-09-01}\ } (\bibinfo {year} {2009})}\BibitemShut {NoStop}%
\bibitem [{\citenamefont {Griffiths}\ \emph {et~al.}(2020)\citenamefont {Griffiths}, \citenamefont {Kleinegesse}, \citenamefont {Saunders}, \citenamefont {Taylor},\ and\ \citenamefont {Vacheret}}]{Griffiths2020}%
  \BibitemOpen
  \bibfield  {author} {\bibinfo {author} {\bibfnamefont {J.}~\bibnamefont {Griffiths}}, \bibinfo {author} {\bibfnamefont {S.}~\bibnamefont {Kleinegesse}}, \bibinfo {author} {\bibfnamefont {D.}~\bibnamefont {Saunders}}, \bibinfo {author} {\bibfnamefont {R.}~\bibnamefont {Taylor}},\ and\ \bibinfo {author} {\bibfnamefont {A.}~\bibnamefont {Vacheret}},\ }\href {https://doi.org/10.1088/2632-2153/abb781} {\bibfield  {journal} {\bibinfo  {journal} {Machine Learning: Science and Technology}\ }\textbf {\bibinfo {volume} {1}},\ \bibinfo {pages} {045022} (\bibinfo {year} {2020})}\BibitemShut {NoStop}%
\bibitem [{\citenamefont {Chen}\ and\ \citenamefont {Guestrin}(2016)}]{xgboost}%
  \BibitemOpen
  \bibfield  {author} {\bibinfo {author} {\bibfnamefont {T.}~\bibnamefont {Chen}}\ and\ \bibinfo {author} {\bibfnamefont {C.}~\bibnamefont {Guestrin}},\ }in\ \href {https://doi.org/10.1145/2939672.2939785} {\emph {\bibinfo {booktitle} {Proceedings of the 22nd ACM SIGKDD International Conference on Knowledge Discovery and Data Mining}}},\ \bibinfo {series and number} {KDD '16}\ (\bibinfo {year} {2016})\ pp.\ \bibinfo {pages} {785--794}\BibitemShut {NoStop}%
\bibitem [{\citenamefont {Galbinski}(2024)}]{Galbinski2024}%
  \BibitemOpen
  \bibfield  {author} {\bibinfo {author} {\bibfnamefont {D.}~\bibnamefont {Galbinski}},\ }\href {https://doi.org/https://doi.org/10.25560/110314} {Ph.D. thesis},\ \bibinfo  {school} {Imperial College London} (\bibinfo {year} {2024})\BibitemShut {NoStop}%
\bibitem [{\citenamefont {Koch}(2019)}]{remu2}%
  \BibitemOpen
  \bibfield  {author} {\bibinfo {author} {\bibfnamefont {L.}~\bibnamefont {Koch}},\ }\href {https://doi.org/10.1088/1748-0221/14/09/P09013} {\bibfield  {journal} {\bibinfo  {journal} {Journal of Instrumentation}\ }\textbf {\bibinfo {volume} {14}}\bibinfo  {number} { (09)},\ \bibinfo {pages} {P09013}}\BibitemShut {NoStop}%
\bibitem [{\citenamefont {Metropolis}\ \emph {et~al.}(1953)\citenamefont {Metropolis}, \citenamefont {Rosenbluth}, \citenamefont {Rosenbluth}, \citenamefont {Teller},\ and\ \citenamefont {Teller}}]{Metropolis1953}%
  \BibitemOpen
\bibfield  {number} {  }\bibfield  {author} {\bibinfo {author} {\bibfnamefont {N.}~\bibnamefont {Metropolis}}, \bibinfo {author} {\bibfnamefont {A.~W.}\ \bibnamefont {Rosenbluth}}, \bibinfo {author} {\bibfnamefont {M.~N.}\ \bibnamefont {Rosenbluth}}, \bibinfo {author} {\bibfnamefont {A.~H.}\ \bibnamefont {Teller}},\ and\ \bibinfo {author} {\bibfnamefont {E.}~\bibnamefont {Teller}},\ }\href {https://doi.org/10.1063/1.1699114} {\bibfield  {journal} {\bibinfo  {journal} {The Journal of Chemical Physics}\ }\textbf {\bibinfo {volume} {21}},\ \bibinfo {pages} {1087} (\bibinfo {year} {1953})}\BibitemShut {NoStop}%
\bibitem [{\citenamefont {Hastings}(1970)}]{Hastings1970}%
  \BibitemOpen
  \bibfield  {author} {\bibinfo {author} {\bibfnamefont {W.~K.}\ \bibnamefont {Hastings}},\ }\href {https://doi.org/10.1093/biomet/57.1.97} {\bibfield  {journal} {\bibinfo  {journal} {Biometrika}\ }\textbf {\bibinfo {volume} {57}},\ \bibinfo {pages} {97} (\bibinfo {year} {1970})}\BibitemShut {NoStop}%
\bibitem [{\citenamefont {Feldman}\ and\ \citenamefont {Cousins}(1998)}]{Feldman1998}%
  \BibitemOpen
  \bibfield  {author} {\bibinfo {author} {\bibfnamefont {G.~J.}\ \bibnamefont {Feldman}}\ and\ \bibinfo {author} {\bibfnamefont {R.~D.}\ \bibnamefont {Cousins}},\ }\href {https://doi.org/10.1103/PhysRevD.57.3873} {\bibfield  {journal} {\bibinfo  {journal} {Physical Review D}\ }\textbf {\bibinfo {volume} {57}},\ \bibinfo {pages} {3873} (\bibinfo {year} {1998})}\BibitemShut {NoStop}%
\end{thebibliography}%

\end{document}